\documentclass[a4paper,fleqn]{cas-sc}
\usepackage[numbers,sort&compress]{natbib}
\usepackage{bm}
\usepackage{mathtools}

\makeatletter
\def\Hy@Warning#1{}
\makeatother

\usepackage{longtable}

\setcitestyle{numbers,open={[},close={]}}

\def\tsc#1{\csdef{#1}{\textsc{\lowercase{#1}}\xspace}}
\tsc{WGM}
\tsc{QE}
\tsc{EP}
\tsc{PMS}
\tsc{BEC}
\tsc{DE}

\begin{document}
\let\WriteBookmarks\relax
\def\floatpagepagefraction{1}
\def\textpagefraction{.001}
\shortauthors{Theisen and Singh}

\title[mode = title]{
Effects of Thermal Boundary Conditions on Natural Convection and Entropy Generation in Non-Newtonian Power-Law Fluids
}

\shorttitle{Thermal Boundary Effects on Non-Newtonian Natural Convection}
\author{Lambert Theisen}[%
orcid=0000-0001-5460-5425%
]
\ead{lambert.theisen@rwth-aachen.de}

\author{Satyvir Singh}[%
	orcid=0000-0001-6669-5296%
]
\cormark[1]
\ead{singh@acom.rwth-aachen.de}
\cortext[cor1]{Corresponding author}
\affiliation{organization={Institute for Applied and Computational Mathematics, RWTH Aachen University},
postcode={52062}, 
state={Aachen},
country={Germany}}


\begin{abstract}
This study investigates the role of thermal boundary conditions on natural convection and entropy generation in non-Newtonian power-law fluids confined within a square cavity and a concentric cylindrical annulus. Steady, two-dimensional governing equations based on the incompressible power-law model and the Boussinesq approximation are solved using the Gridap.jl finite element framework. The numerical methodology is validated against benchmark solutions for both Newtonian and non-Newtonian convection, showing good agreement in terms of isotherm fields, streamlines, local Nusselt number distributions, and entropy generation. The effects of fluid rheology and heating mode are examined for shear-thinning, Newtonian, and shear-thickening fluids under uniform and non-uniform thermal boundary conditions. The results show that shear-thinning behavior enhances buoyancy-driven circulation, steepens thermal gradients, and increases heat transfer, whereas shear-thickening behavior suppresses convection and promotes conduction-dominated transport. Thermal boundary conditions are found to play an important role in controlling the intensity and spatial distribution of flow, heat transfer, and irreversibility. In both geometries, uniform heating produces stronger and more distributed convective structures, while non-uniform sinusoidal heating localizes thermal forcing and consistently reduces total entropy generation. An entropy analysis further reveals that viscous dissipation dominates irreversibility in shear-thinning fluids, whereas heat-transfer irreversibility becomes dominant as the power-law index increases. The study demonstrates that appropriate thermal boundary design, together with fluid rheology, provides an effective route for controlling heat transfer and minimizing thermodynamic losses in non-Newtonian convection systems. The source code and metadata are publicly available.
\end{abstract}


\begingroup
	\hfuzz=200pt 
	\maketitle
\endgroup

\section{Introduction}
\label{Sec:1}

Natural convection, driven by buoyancy forces arising from temperature-induced density differences, plays a fundamental role in a wide range of engineering, industrial, and environmental processes, including cooling of electronic equipment, building thermal systems, solar energy applications, and large-scale industrial flows~\cite{incropera1988convection, al2006analysis, ali2023numerical, pakdaman2011performance, novev2018natural}. The phenomenon occurs when temperature differences near a heated or cooled surface change the fluid's density. In a gravitational field, the resulting buoyancy forces cause warmer, less dense fluid to rise and cooler, denser fluid to sink, producing flow without external mechanical forcing such as a pump or fan. This passive mode of heat transfer is energy-efficient and has been extensively studied over several decades in both fundamental and applied contexts~\cite{bejan2013convection, calmidi2000forced, buchberg1976natural, theisen2023simulation}. Despite its apparent simplicity, natural convection involves complex interactions among momentum, energy, and mass conservation mechanisms, particularly when the working fluid exhibits non-Newtonian behavior~\cite{guha2014natural, yang2020comprehensive, shenoy1982thermal, khanafer2003buoyancy}.

A wide range of industrial and biological fluids, including polymer solutions, molten plastics, paints, blood, and drilling muds, exhibit non-Newtonian rheological behavior that significantly modifies convective flow dynamics and associated heat transfer characteristics~\cite{bird1987dynamics, chhabra2010non, tanner2000engineering}. Among the various constitutive models developed to describe such fluids, the power-law (Ostwald--de Waele) model has gained particular prominence due to its simplicity and its ability to represent both shear-thinning ($n<1$) and shear-thickening ($n>1$) behaviors through the power-law index $n$~\cite{metzner1955flow, yang2020comprehensive, hatami2014natural}. In this framework, the apparent viscosity depends on the local shear rate as $\mu_{\text{eff}} = K \dot{\gamma}^{\,n-1}$, where $K$ denotes the consistency index. The interaction between shear-dependent viscosity and buoyancy-driven momentum transport leads to complex flow structures that differ markedly from those observed in Newtonian fluids, influencing thermal boundary layers and heat transfer rates. Consequently, classical correlations for Newtonian convection are no longer directly applicable, and dedicated analyses are required~\cite{turan2011laminar, khezzar2012natural, safaei2011numerical, metzner1965heat}. Understanding natural convection in power-law fluids is therefore of both fundamental scientific interest and practical importance for the design and optimization of thermal systems involving complex fluids.


Thermal boundary conditions play an important role in determining the flow structure, temperature distribution, and heat transfer characteristics in convective systems. In natural convection problems, the way in which thermal energy is supplied or removed at the boundaries fundamentally governs the buoyancy forces that drive the fluid motion. The most commonly employed thermal boundary conditions in the literature include the isothermal condition, where a uniform constant temperature is maintained at the wall, and the isoflux condition, where a uniform heat flux is prescribed at the boundary~\cite{basak2006effects, basak2006natural}. Additionally, more complex conditions such as linearly varying wall temperature, sinusoidal temperature distribution, and mixed or partial heating have been investigated to simulate realistic engineering scenarios~\cite{roy2005finite, sathiyamoorthy2007steady, nouri2024non}. Each of these boundary conditions produces different thermal gradients within the fluid domain, resulting in qualitatively and quantitatively different velocity and temperature fields, as well as varying rates of convective heat transfer expressed through the local and average Nusselt number~\cite{basak2006effects, roy2005finite, pan2025thermal}.

The choice of thermal boundary condition is not merely a mathematical convenience but reflects the physical reality of the application under consideration. For instance, isothermal boundaries are representative of condensing or evaporating surfaces and highly conductive walls, whereas constant heat flux conditions are more appropriate for electrically heated surfaces, solar collectors, and nuclear reactor cooling channels~\cite{basak2006natural, basak2007finite}. In the context of non-Newtonian power-law fluids, the sensitivity to thermal boundary conditions is further amplified because the effective viscosity depends on the local shear rate, which itself is shaped by the temperature-driven velocity gradients near the wall~\cite{basak2007finite, m2024natural}. This two-way coupling between the thermal field and the rheological response makes the selection and analysis of thermal boundary conditions a particularly critical aspect of non-Newtonian convection studies~\cite{basak2008role, rehman2024multigrid}. Consequently, a systematic comparison of different thermal boundary conditions is essential to develop a comprehensive understanding of heat transfer behavior in power-law fluid systems and to provide reliable guidelines for practical thermal engineering design~\cite{basak2011effects, mahmood2025thermo}.


The second law of thermodynamics provides an additional framework for evaluating 
the performance of thermal systems beyond what is possible through energy 
conservation alone. Entropy generation, which quantifies the irreversibility 
associated with heat transfer, viscous dissipation, and mass transfer within 
a system, serves as a direct measure of the thermodynamic inefficiency of a 
convective process~\cite{bejan2013entropy}. The concept was first introduced by Adrian Bejan~\cite{bejan1979study}, who showed that minimizing entropy generation provides a fundamental basis for the design of thermally optimal systems. In buoyancy-driven 
convective flows, entropy is generated due to two primary mechanisms: 
irreversibility arising from finite temperature gradients, referred to as 
thermal entropy generation, and irreversibility due to fluid friction, 
referred to as viscous entropy generation~\cite{bejan2013entropy, 
	oztop2012entropy}. The relative dominance of these two contributions is 
conveniently characterized by the Bejan number, defined as the ratio of 
thermal entropy generation to total entropy generation, which approaches 
unity when heat transfer irreversibility dominates and tends toward zero 
when viscous dissipation effects prevail~\cite{bejan1979study, 
	oztop2012entropy}. A thorough understanding of entropy generation and its 
spatial distribution within the flow domain is therefore essential for 
identifying thermodynamically inefficient regions and guiding the design 
of energy-optimal thermal systems.

In the context of non-Newtonian power-law fluids, entropy generation 
analysis acquires additional complexity due to the nonlinear dependence 
of the effective viscosity on the local shear rate. Unlike Newtonian 
fluids where viscous dissipation is governed by a constant dynamic 
viscosity, the spatially varying effective viscosity in power-law fluids 
introduces a strong coupling between the flow kinematics and the local 
entropy production rate~\cite{saouli2004second, mahmud2002second}. For 
shear-thinning fluids, the reduced effective viscosity in high-shear 
regions tends to suppress viscous entropy generation while simultaneously 
enhancing convective heat transfer, whereas shear-thickening fluids 
exhibit the opposite trend, with elevated viscosity amplifying frictional 
irreversibilities~\cite{saouli2004second}. Furthermore, the thermal 
boundary conditions imposed on the enclosure walls directly govern the 
temperature gradients within the fluid domain, thereby modulating the 
magnitude and spatial distribution of thermal entropy generation~\cite{basak2011effects, oztop2012entropy}. Consequently, a systematic 
entropy generation analysis that accounts for the combined effects of 
power-law rheology and thermal boundary conditions is indispensable for 
a complete thermodynamic characterization of non-Newtonian natural 
convection systems, and this interplay remains an open and active area 
of research~\cite{basak2011effects, mahmud2002second}.

The study of natural convection in non-Newtonian fluids has experienced remarkable growth over the past few decades, driven by the increasing demand for accurate thermal modeling in polymer processing, food engineering, and biomedical applications. Early investigations by Acrivos et al.~\cite{acrivos1960} established the theoretical foundations of boundary layer natural convection for power-law fluids past heated surfaces, demonstrating that the power-law index significantly alters the velocity and temperature boundary layer thicknesses. Subsequent numerical and analytical studies extended these findings to enclosed cavities, with significant contributions from Ozoe and Churchill~\cite{ozoe1972}, who examined the influence of rheological behavior on heat transfer rates in differentially heated enclosures. A major advancement was made by Turan et al.~\cite{turan2010}, who performed systematic two-dimensional steady-state simulations of laminar natural convection of power-law fluids in square enclosures with differentially heated sidewalls subjected to constant wall temperatures, providing detailed correlations between the mean Nusselt number, Rayleigh number, Prandtl number, and power-law index. The same group further extended their analysis to constant wall heat flux boundary conditions~\cite{turan2012}, demonstrating that the choice of thermal boundary condition leads to quantitatively distinct heat transfer scaling behaviors, particularly for strongly shear-thinning fluids. These foundational contributions collectively confirmed that non-Newtonian rheology introduces qualitatively distinct flow patterns and heat transfer mechanisms compared to classical Newtonian natural convection, necessitating dedicated computational frameworks tailored to power-law constitutive behavior~\cite{lamsaadi2006, kim2003}.

More recent investigations have increasingly focused on the combined role of thermal boundary conditions, enclosure geometry, and rheological nonlinearity in governing the convective behavior of power-law fluids. Lamsaadi et al.~\cite{lamsaadi2006} examined natural convection of power-law fluids in a shallow horizontal rectangular cavity subjected to uniform heat flux boundary conditions, demonstrating that shear-thinning fluids exhibit enhanced heat transfer under isoflux conditions compared to isothermal heating. The influence of sinusoidal and non-uniform thermal boundary conditions on power-law fluid convection in square and trapezoidal cavities has been explored by Basak and co-workers~\cite{basak2007finite, sathiyamoorthy2007steady}, with results indicating that the spatial periodicity of the wall temperature distribution leads to multiple convective cell formations and non-monotonic Nusselt number variations. More recently, numerical studies on power-law nanofluids in square cavities with internal fins under different thermal boundary conditions~\cite{m2024natural} confirmed that the Nusselt number enhancement is strongly amplified for shear-thinning nanofluids, with improvements exceeding 130\% compared to Newtonian behavior at high Rayleigh numbers. Further contributions using multigrid solvers~\cite{rehman2024multigrid} have reinforced the sensitivity of flow structures and thermal irreversibilities to the imposed boundary conditions in power-law fluid systems. Despite these advances, a systematic and comprehensive comparative study that simultaneously examines the effects of multiple thermal boundary conditions on both the convective heat transfer and the entropy generation characteristics of power-law fluids remains scarce in the open literature~\cite{basak2011effects, oztop2012entropy}, thereby motivating the present work.

Despite the considerable body of literature on natural convection in 
power-law fluids, a critical examination reveals several important gaps 
that remain unaddressed. The majority of earlier investigations have 
treated flow behavior, heat transfer, and entropy generation as largely 
separate problems, with very few studies considering these quantities 
within a unified analysis framework~\cite{oztop2012entropy, basak2011effects}. While the influence of 
thermal boundary conditions on Newtonian natural convection has been 
extensively documented~\cite{basak2006effects, basak2006natural, 
	roy2005finite}, the corresponding systematic analysis for non-Newtonian 
power-law fluids subjected to multiple distinct thermal boundary 
conditions remains largely absent~\cite{turan2010, turan2012}. Moreover, 
entropy generation studies for power-law fluids that simultaneously 
account for rheological nonlinearity, power-law index effects, and 
the type of imposed thermal boundary condition are scarce in the open 
literature~\cite{saouli2004second, mahmud2002second, basak2011effects}. 

To address these gaps, the specific objectives of the present 
work are: (i) to develop and validate a robust finite element solver for 
steady-state laminar natural convection of power-law fluids in a square 
cavity and a concentric cylindrical annulus~\cite{turan2010, 
	basak2006effects}; (ii) to investigate the combined influence of the 
power-law index \(n\) and thermal boundary conditions, including uniform 
and non-uniform heating, on the flow structure and heat transfer 
characteristics expressed through local and average Nusselt numbers~\cite{turan2010, turan2012, basak2006natural}; (iii) to evaluate the 
spatial distribution of local entropy generation due to heat transfer 
and viscous dissipation, total entropy generation, and average Bejan 
number across a range of power-law indices and thermal boundary 
conditions~\cite{bejan1979study, bejan2013entropy, basak2011effects}; 
and (iv) to provide physical insight into the interplay between 
shear-thinning \((n < 1)\), Newtonian \((n = 1)\), and shear-thickening 
\((n > 1)\) behaviors and the imposed thermal boundary conditions on 
heat-transfer performance and thermodynamic irreversibility~\cite{saouli2004second, oztop2012entropy}. The findings of this study 
are expected to contribute to the optimization of energy-efficient 
thermal systems involving complex fluids and diverse enclosure designs.

The remainder of this manuscript is organized as follows. 
Section~\ref{Sec:2} presents the mathematical formulation, including 
the governing equations, thermal boundary conditions, and power-law 
constitutive model. Section~\ref{Sec:3} describes the finite element 
framework and nonlinear solution strategy. Then, Section~\ref{Sec:4} 
introduces the post-processing quantities, including the stream 
function, heat function, Nusselt number, entropy generation, and Bejan 
number. Validation studies 
for both Newtonian and non-Newtonian natural convection, benchmarked 
against established results from the literature, are presented in 
Section~\ref{Sec:5}. The effects of the power-law index and 
thermal boundary conditions on flow structure, heat transfer, and 
entropy generation in both geometries are discussed in 
Section~\ref{Sec:6}. Finally, the key conclusions and outlook of this study are 
summarized in Section~\ref{Sec:7}.

\begin{figure}
	\centering
	\includegraphics[width=0.8\textwidth]{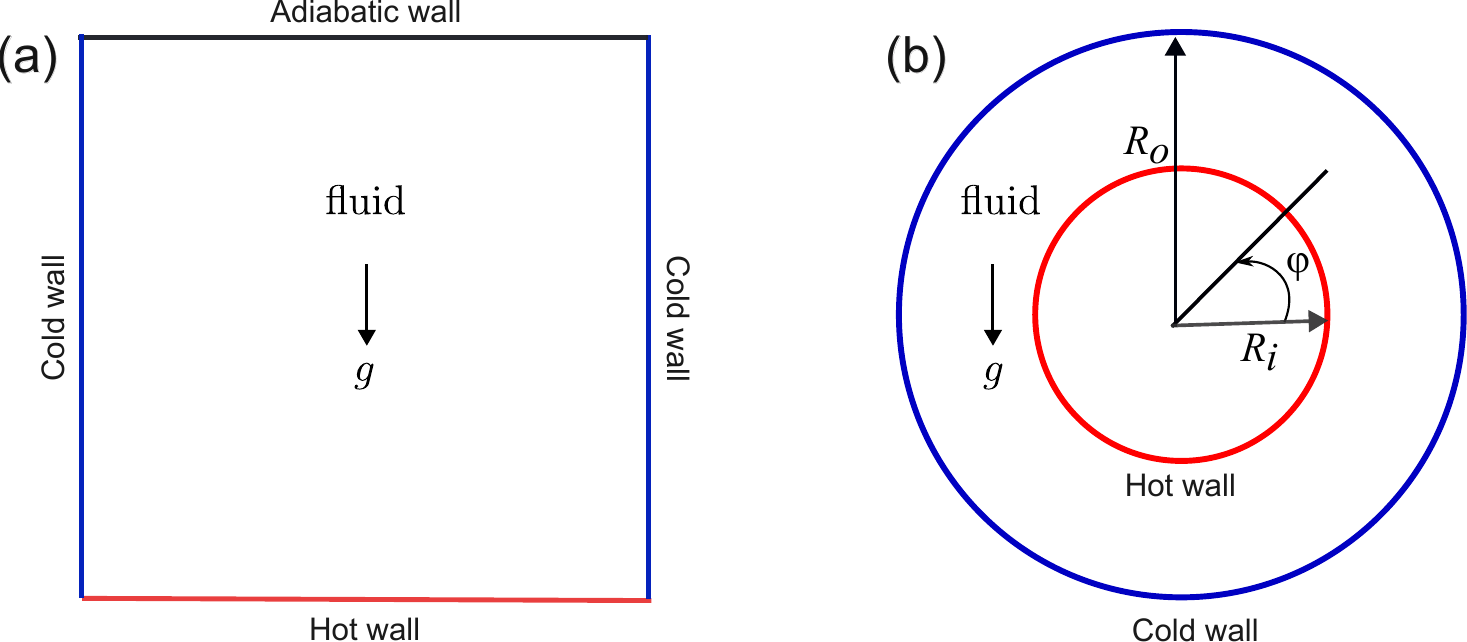}
	\caption{Schematic diagrams of the physical domains considered in the study of natural convection and entropy generation in power-law fluids.\ (a) Square cavity with a hot bottom wall, cold vertical side walls, and an adiabatic top wall.\ (b) Concentric cylindrical annulus with an inner hot cylinder of radius $R_i$ and an outer cold cylinder of radius \(R_o\), where \(\varphi\) denotes the angular coordinate. In both configurations, the gravitational acceleration \(g\) acts vertically downward.}\label{Fig:1}  
\end{figure}

\section{Mathematical formulation} \label{Sec:2}

\subsection{Problem Setup}\label{Sec:2.1}

This study investigates steady-state natural convection and entropy 
generation in non-Newtonian power-law fluids confined within two 
distinct two-dimensional geometrical configurations, including a square cavity and a concentric cylindrical annulus, as illustrated 
in Figure~\ref{Fig:1}. In both configurations, the flow is driven by 
buoyancy forces arising from temperature differences imposed at the 
domain boundaries, and the gravitational acceleration $g$ acts 
vertically downward. In the square cavity configuration, shown in Fig.~\ref{Fig:1}(a), the computational domain is a unit square filled with a power-law fluid. The bottom wall is subjected to either uniform heating or non-uniform sinusoidal heating, representing the hot wall, while the top 
wall is thermally insulated, i.e., adiabatic with zero heat flux. The two vertical side walls  are maintained at a constant lower temperature, serving as the cold walls. The imposed temperature difference between the 
hot bottom wall and the cold side walls drives the buoyancy-induced 
fluid motion within the enclosure. In the concentric cylindrical annulus configuration, shown in  Fig.~\ref{Fig:1}(b), the fluid occupies the annular region between two coaxial cylinders. The inner cylinder of radius $R_{i}$ is 
subjected to either uniform heating or non-uniform sinusoidal heating, while the 
outer cylinder of radius $R_{o}$ is kept at a lower constant 
temperature (cold wall), with the characteristic length defined as 
the gap width $L = R_{o} - R_{i}$. The angular coordinate $\varphi$ 
describes the circumferential direction of flow within the annular 
domain. Natural convection patterns are driven by the radial 
temperature gradient between the inner hot and outer cold cylinders, 
with gravity acting vertically downward.

\subsection{Governing equations}\label{Sec:2.0}

The present study investigates steady-state, two-dimensional, laminar natural convection of an incompressible non-Newtonian power-law fluid confined within a square cavity and a concentric cylindrical annulus. The flow is driven by buoyancy forces induced by temperature gradients imposed through thermal boundary conditions. The governing equations for the flow and thermal fields are formulated based on the conservation of mass, momentum, and energy under the assumptions that the fluid is incompressible with constant thermophysical properties and that the Boussinesq approximation is valid, whereby density variations are considered only in the buoyancy term while neglected elsewhere. Furthermore, thermal radiation is neglected, and viscous heating is omitted from the energy equation; the associated fluid-friction irreversibility is evaluated separately in the entropy-generation post-processing. Under these assumptions, the dimensional governing equations in Cartesian coordinates $(x,y)$ are given as follows~\cite{basak2006effects},
\begin{align}
	\frac{\partial u}{\partial x} + \frac{\partial v}{\partial y} 
	&= 0,
	\label{eq:1} \\
	\rho\left(u\frac{\partial u}{\partial x} + v\frac{\partial u}
	{\partial y}\right) &= -\frac{\partial p}{\partial x} + 
	\frac{\partial \tau_{xx}}{\partial x} + 
	\frac{\partial \tau_{xy}}{\partial y},
	\label{eq:2} \\
	\rho\left(u\frac{\partial v}{\partial x} + v\frac{\partial v}
	{\partial y}\right) &= -\frac{\partial p}{\partial y} + 
	\frac{\partial \tau_{xy}}{\partial x} + 
	\frac{\partial \tau_{yy}}{\partial y} + \rho g\beta(T - T_c),
	\label{eq:3} \\
	\rho c_p\left(u\frac{\partial T}{\partial x} + v
	\frac{\partial T}{\partial y}\right) &= \kappa\left(
	\frac{\partial^2 T}{\partial x^2} + 
	\frac{\partial^2 T}{\partial y^2}\right),
	\label{eq:4}
\end{align}
where $(u,v)$ are the velocity components, $p$ is the pressure, 
$T$ is the temperature, $\rho$ is the fluid density, $g$ is the 
gravitational acceleration, $\beta$ is the thermal expansion 
coefficient, $\kappa$ is the thermal conductivity, $c_p$ is the 
specific heat capacity, and $\tau_{ij}$ is the shear stress tensor.

\subsection{Power-Law Constitutive Model}\label{Sec:2.3}

The rheological behavior of the working fluid is described by the 
power-law (Ostwald--de Waele) constitutive equation~\cite{matin2013natural}. The shear 
stress tensor $\boldsymbol{\tau} \in \mathbb{R}^{2\times 2}$ is 
defined as 
\begin{equation}
	\tau_{ij} = 2\mu_a\, D_{ij}, \qquad 
	D_{ij} = \frac{1}{2}\left(\frac{\partial u_i}{\partial x_j} + 
	\frac{\partial u_j}{\partial x_i}\right),
	\label{eq:5}
\end{equation}
where $D_{ij}$ are the components of the rate-of-deformation tensor $\boldsymbol{D}$, and $\mu_a$ is the apparent dynamic 
viscosity defined as
\begin{equation}
	\mu_a = K\left(2\boldsymbol{D}:\boldsymbol{D}\right)^{\frac{n-1}
		{2}} = K\left\{2\left[\left(\frac{\partial u}{\partial x}
	\right)^2 + \left(\frac{\partial v}{\partial y}\right)^2\right] 
	+ \left(\frac{\partial u}{\partial y} + \frac{\partial v}
	{\partial x}\right)^2\right\}^{\frac{n-1}{2}},
	\label{eq:6}
\end{equation}
where $K > 0$ is the consistency coefficient and $n > 0$ is the 
power-law index. The model recovers a Newtonian fluid with constant 
viscosity $\mu = K$ for $n = 1$.

\subsection{Non-Dimensionalization}\label{Sec:2.4}

The governing equations are cast in dimensionless form using the 
following reference scales. The cavity side length (or gap width 
for the annulus) $L$ is chosen as the length scale, the thermal 
diffusivity-based velocity $\alpha/L$ as the velocity scale, $T_{h}$ and $T_{c}$ are the temperatures at hot and cold walls, respectively, and 
$\rho(\alpha/L)^2$ as the pressure scale~\cite{turan2011laminar,matin2013laminar}.
\begin{equation}
	X = \frac{x}{L},\quad Y = \frac{y}{L},\quad 
	U = \frac{uL}{\alpha},\quad V = \frac{vL}{\alpha},\quad 
	\theta = \frac{T - T_c}{T_h - T_c},\quad 
	P = \frac{pL^2}{\rho\alpha^2}.
	\label{eq:7}
\end{equation}
The apparent viscosity is scaled by the shear rate associated with the 
diffusive velocity scale, such that 
$\mu_a = K(\alpha/L^2)^{n-1}\bar{\mu}$. The resulting dimensionless 
apparent viscosity is
\begin{equation}
	\bar{\mu} = \left\{2\left[\left(
	\frac{\partial U}{\partial X}\right)^2 + \left(
	\frac{\partial V}{\partial Y}\right)^2\right] + 
	\left(\frac{\partial U}{\partial Y} + 
	\frac{\partial V}{\partial X}\right)^2
	\right\}^{\frac{n-1}{2}},
	\label{eq:8}
\end{equation}
and the modified Prandtl and Rayleigh numbers for power-law fluids are 
defined following Ng and Hartnett's convention as used by Turan et 
al.~\cite{turan2011laminar} and Matin and Khan~\cite{matin2013laminar}:
\begin{equation}
	\operatorname{Pr} = \frac{K}{\rho}\,
	\frac{L^{2-2n}}{\alpha^{2-n}}, \qquad 
	\operatorname{Ra} = 
	\frac{g\beta\Delta T\, L^{2n+1}}
	{(K/\rho)\,\alpha^n},
	\label{eq:9}
\end{equation}
where $\alpha = \kappa/(\rho c_p)$ is the thermal diffusivity and 
$\Delta T = T_h - T_c$. For $n=1$, these definitions reduce to the 
classical Newtonian forms $\operatorname{Pr}=\nu/\alpha$ and 
$\operatorname{Ra}=g\beta\Delta T L^3/(\nu\alpha)$ with 
$\nu=K/\rho$. Substituting 
Eqs.~\eqref{eq:7}--\eqref{eq:9} into 
Eqs.~\eqref{eq:1}--\eqref{eq:4}, the dimensionless 
governing equations become
\begin{align}
	\frac{\partial U}{\partial X} + \frac{\partial V}{\partial Y} 
	&= 0,
	\label{eq:10} \\
	U\frac{\partial U}{\partial X} + V\frac{\partial U}{\partial Y} 
	&= -\frac{\partial P}{\partial X} + \operatorname{Pr}\left[
	2\frac{\partial}{\partial X}\!\left(\bar{\mu}
	\frac{\partial U}{\partial X}\right) + 
	\frac{\partial}{\partial Y}\!\left(\bar{\mu}\left(
	\frac{\partial U}{\partial Y} + 
	\frac{\partial V}{\partial X}\right)\right)\right],
	\label{eq:11} \\
	U\frac{\partial V}{\partial X} + V\frac{\partial V}{\partial Y} 
	&= -\frac{\partial P}{\partial Y} + \operatorname{Pr}\left[
	\frac{\partial}{\partial X}\!\left(\bar{\mu}\left(
	\frac{\partial U}{\partial Y} + 
	\frac{\partial V}{\partial X}\right)\right) + 
	2\frac{\partial}{\partial Y}\!\left(\bar{\mu}
	\frac{\partial V}{\partial Y}\right)\right] 
	+ \operatorname{Ra}\operatorname{Pr}\theta,
	\label{eq:12} \\
	U\frac{\partial\theta}{\partial X} + 
	V\frac{\partial\theta}{\partial Y} &= 
	\frac{\partial^2\theta}{\partial X^2} + 
	\frac{\partial^2\theta}{\partial Y^2}.
	\label{eq:13}
\end{align}
In vector notation, the buoyancy term appearing in the vertical momentum 
equation is written as 
$\operatorname{Ra}\operatorname{Pr}\theta\,\widehat{\boldsymbol{g}}$, 
where $\widehat{\boldsymbol{g}}=(0,1)^\top$ denotes the dimensionless 
buoyancy direction, opposite to the downward gravitational acceleration.

\subsection{Boundary Conditions}\label{Sec:2.5}

The dimensionless boundary conditions corresponding to Eqs.~\eqref{eq:10}--\eqref{eq:13} are prescribed according to the geometry under consideration. For the square cavity, the boundary conditions are given by
\begin{equation}
	\left\{
	\begin{aligned}
		&U = V = 0 \quad \text{on all walls (no-slip),}\\
		&\theta(X,0) = 1 \quad \text{(uniform heating)} \quad 
		\text{or} \quad \theta(X,0) = \sin(\pi X) \quad 
		\text{(non-uniform heating),}\\
		&\theta(0,Y) = \theta(1,Y) = 0 \quad \text{(cold side walls),}\\
		&\frac{\partial\theta}{\partial Y}(X,1) = 0 \quad 
		\text{(adiabatic top wall).}
	\end{aligned}
	\right.
	\label{eq:14}
\end{equation}
For the concentric cylindrical annulus, the corresponding boundary conditions are expressed as
\begin{equation}
	\left\{
	\begin{aligned}
		&U = V = 0,\quad \text{on all walls (no-slip),}\\
		&\theta = 0 \quad \text{at } r = R_o 
		\quad\text{(cold outer wall),}\\
		&U = V = 0,\quad \theta = 1 \quad \text{(uniform heating)} 
		\quad \text{or}\\
		&\theta = 0.5 (1 +  \sin\varphi) \quad 
		\text{(non-uniform heating)} \quad 
		\text{at } r = R_i \quad\text{(hot inner wall).}
	\end{aligned}
	\right.
	\label{eq:15}
\end{equation}

\section{Finite element framework and numerical solution}\label{Sec:3}

To obtain a numerical solution of the dimensionless governing 
Eqs.~\eqref{eq:10}--\eqref{eq:13} subject to the 
boundary conditions~\eqref{eq:14}--\eqref{eq:15}, the 
Galerkin finite element method (FEM) is employed. The implementation 
is carried out using the \texttt{Gridap.jl} finite element 
framework~\cite{badiaGridapExtensibleFinite2020, 
	verdugoSoftwareDesignGridap2022} within the Julia programming 
language~\cite{bezansonJuliaFreshApproach2017}. The computational 
domain $\Omega$ is partitioned into a finite number of 
non-overlapping elements, and the continuous solution fields are 
approximated using polynomial basis functions within each element. 
The full derivation of the weak formulation is provided in 
Appendix~\ref{Appendix:B}. For reproducibility, the source code and all metadata are available at~\cite{theisen2026gemotion}.

\subsection{Element Choice and Discrete Approximation}
\label{Sec:3.1}

Taylor--Hood finite elements are employed, which consist of 
continuous second-order Lagrange polynomials
for the velocity components $(U, V)$ and temperature $\theta$, 
and continuous first-order Lagrange polynomials
for the pressure $P$. The pressure space is constrained to have zero 
mean over the computational domain, which fixes the pressure null space 
associated with the incompressible formulation. This inf-sup stable element pair ensures a 
well-posed discrete velocity-pressure coupling and avoids spurious 
pressure modes. The discrete solution fields are expressed as 
linear combinations of nodal basis functions
\begin{equation}
	U_h(\boldsymbol{x}) = \sum_{i=1}^{N_u} U_i\,\Phi_i^u(\boldsymbol{x}),
	\quad
	V_h(\boldsymbol{x}) = \sum_{i=1}^{N_u} V_i\,\Phi_i^u(\boldsymbol{x}),
	\quad
	P_h(\boldsymbol{x}) = \sum_{i=1}^{N_p} P_i\,\Phi_i^p(\boldsymbol{x}),
	\quad
	\theta_h(\boldsymbol{x}) = \sum_{i=1}^{N_\theta} 
	\theta_i\,\Phi_i^\theta(\boldsymbol{x}),
	\label{eq:16}
\end{equation}
where $U_i$, $V_i$, $P_i$, and $\theta_i$ are the nodal degrees 
of freedom (DOFs); $\Phi_i^u$, $\Phi_i^p$, and $\Phi_i^\theta$ 
are the basis functions for velocity, pressure, and temperature, 
respectively; and $N_u$, $N_p$, and $N_\theta$ denote the total 
number of velocity, pressure, and temperature DOFs. Dirichlet boundary 
conditions are imposed through affine trial spaces for velocity and 
temperature, whereas the corresponding test spaces satisfy homogeneous 
Dirichlet conditions. In the Galerkin method, the test functions use the 
same polynomial orders as the trial functions, and testing with the basis 
functions leads to a system of algebraic equations for the nodal DOFs as detailed in the next section.
%

\subsection{Discrete Residual Equations}
\label{Sec:3.2}

Substituting the discrete approximations~\eqref{eq:16}
into the weak formulation (see Appendix~\ref{Appendix:B}) yields the discrete nonlinear residual vector
$\mathbf{R} = (\mathbf{R}_U,\, \mathbf{R}_V,\, \mathbf{R}_P,\, 
\mathbf{R}_\theta)^\top$, whose components are computed as follows
%
\begin{equation}
	\begin{aligned}
		R_{U,i} &= \sum_{j=1}^{N_u} U_j \int_\Omega \left(U_h 
		\frac{\partial\Phi_j^u}{\partial X} + V_h 
		\frac{\partial\Phi_j^u}{\partial Y}\right)\Phi_i^u\,
		\mathrm{d}\Omega 
		- \sum_{j=1}^{N_p} P_j \int_\Omega \Phi_j^p 
		\frac{\partial\Phi_i^u}{\partial X}\,\mathrm{d}\Omega \\
			&+ 2Pr\sum_{j=1}^{N_u} U_j \int_\Omega \bar{\mu}_{\mathrm{s}}\,
			\frac{\partial\Phi_j^u}{\partial X}
			\frac{\partial\Phi_i^u}{\partial X}\,\mathrm{d}\Omega
		  + Pr\sum_{j=1}^{N_u} \int_\Omega \bar{\mu}_{\mathrm{s}}\left(
			U_j\frac{\partial\Phi_j^u}{\partial Y} +
			V_j\frac{\partial\Phi_j^u}{\partial X}\right)
			\frac{\partial\Phi_i^u}{\partial Y}\,\mathrm{d}\Omega,
	\end{aligned}
	\label{eq:18}
\end{equation}
%
\begin{equation}
	\begin{aligned}
		R_{V,i} &= \sum_{j=1}^{N_u} V_j \int_\Omega \left(U_h 
		\frac{\partial\Phi_j^u}{\partial X} + V_h 
		\frac{\partial\Phi_j^u}{\partial Y}\right)\Phi_i^u\,
		\mathrm{d}\Omega 
		- \sum_{j=1}^{N_p} P_j \int_\Omega \Phi_j^p 
		\frac{\partial\Phi_i^u}{\partial Y}\,\mathrm{d}\Omega \\
			&+ Pr\sum_{j=1}^{N_u} \int_\Omega \bar{\mu}_{\mathrm{s}}\left(
			U_j\frac{\partial\Phi_j^u}{\partial Y} +
			V_j\frac{\partial\Phi_j^u}{\partial X}\right)
			\frac{\partial\Phi_i^u}{\partial X}\,\mathrm{d}\Omega
			+ 2Pr\sum_{j=1}^{N_u} V_j \int_\Omega \bar{\mu}_{\mathrm{s}}\,
			\frac{\partial\Phi_j^u}{\partial Y}
			\frac{\partial\Phi_i^u}{\partial Y}\,\mathrm{d}\Omega \\
		&- Ra\,Pr\sum_{j=1}^{N_\theta} \theta_j \int_\Omega 
		\Phi_j^\theta\,\Phi_i^u\,\mathrm{d}\Omega,
	\end{aligned}
	\label{eq:19}
\end{equation}
%
\begin{equation}
	R_{P,i} = \sum_{j=1}^{N_u} U_j \int_\Omega 
	\frac{\partial\Phi_j^u}{\partial X}\Phi_i^p\,\mathrm{d}\Omega 
	+ \sum_{j=1}^{N_u} V_j \int_\Omega 
	\frac{\partial\Phi_j^u}{\partial Y}\Phi_i^p\,\mathrm{d}\Omega,
	\label{eq:20}
\end{equation}
%
\begin{equation}
	R_{\theta,i} = \sum_{j=1}^{N_\theta} \theta_j \int_\Omega 
	\left(U_h\frac{\partial\Phi_j^\theta}{\partial X} + 
	V_h\frac{\partial\Phi_j^\theta}{\partial Y}\right)
	\Phi_i^\theta\,\mathrm{d}\Omega 
	+ \sum_{j=1}^{N_\theta} \theta_j \int_\Omega \left(
	\frac{\partial\Phi_j^\theta}{\partial X}
	\frac{\partial\Phi_i^\theta}{\partial X} + 
	\frac{\partial\Phi_j^\theta}{\partial Y}
	\frac{\partial\Phi_i^\theta}{\partial Y}\right)
	\mathrm{d}\Omega.
	\label{eq:21}
\end{equation}
Note that in Eqs.~\eqref{eq:18}--\eqref{eq:21}, $U_h$ and $V_h$ are the discrete velocity fields from~\eqref{eq:16} evaluated at the
current iterate, which enter the convective terms nonlinearly. The 
viscosity term $\bar{\mu}_{\mathrm{s}}$ also depends nonlinearly on the discrete
velocity field through the stabilized form of Eq.~\eqref{eq:8}, as detailed in Appendix~\ref{Appendix:B}, making the
overall system highly nonlinear, particularly for strongly 
shear-thinning $(n \ll 1)$ or shear-thickening $(n \gg 1)$ fluids. The resulting system of nonlinear algebraic equations is solved using either the Newton--Raphson method or the trust-region method implemented in the 
\texttt{NLsolve.jl} package~\cite{mogensenJuliaNLSolversNLsolvejlV4512020b}, as described in the following sections.

\subsection{Newton--Raphson Solution Method}
\label{Sec:3.3}

The discrete nonlinear system $\mathbf{R}(\mathbf{U}_h) = \mathbf{0}$ 
is solved using the Newton--Raphson method. The global DOF vector is
\begin{equation}
	\mathbf{U}_h = \left(U_1,\ldots,U_{N_u},\; 
	V_1,\ldots,V_{N_u},\; P_1,\ldots,P_{N_p},\; 
	\theta_1,\ldots,\theta_{N_\theta}\right)^\top 
	\in \mathbb{R}^{N_{\mathrm{dof}}},
	\label{eq:22}
\end{equation}
where $N_{\mathrm{dof}} = 2N_u + N_p + N_\theta$ is the total 
number of degrees of freedom. After the prescribed Dirichlet data have 
been incorporated into the affine trial spaces, the free unknowns are 
initialized by a zero vector. At each Newton iteration 
$k = 0, 1, 2, \ldots$, the linearized system
\begin{equation}
	\mathbf{J}^{(k)}\,\delta\mathbf{U}_h^{(k)} = 
	-\mathbf{R}^{(k)},
	\label{eq:23}
\end{equation}
is solved for the update $\delta\mathbf{U}_h^{(k)}$, where 
$\mathbf{J}^{(k)} = (\partial\mathbf{R}/\partial\mathbf{U}_h)
|_{\mathbf{U}_h^{(k)}}$ is the discrete Jacobian matrix
evaluated at the current iterate $\mathbf{U}_h^{(k)}$ and $\mathbf{R}^{(k)}$ is the residual at the same iterate. The Jacobian
has the block structure
		\begin{equation}
			\mathbf{J}^{(k)} =
			\begin{bmatrix}
			\mathbf{J}_{UU}^{(k)} & \mathbf{J}_{UV}^{(k)} & \mathbf{J}_{UP}^{(k)} &
			\mathbf{0} \\[4pt]
			\mathbf{J}_{VU}^{(k)} & \mathbf{J}_{VV}^{(k)} & \mathbf{J}_{VP}^{(k)} &
			\mathbf{J}_{V\theta}^{(k)} \\[4pt]
			\mathbf{J}_{PU}^{(k)} & \mathbf{J}_{PV}^{(k)} & \mathbf{0}       &
			\mathbf{0}        \\[4pt]
			\mathbf{J}_{\theta U}^{(k)} & \mathbf{J}_{\theta V}^{(k)} &
			\mathbf{0} & \mathbf{J}_{\theta\theta}^{(k)}
		\end{bmatrix},
		\label{eq:24}
	\end{equation}
where each block represents the coupling between field
variables $A$ and $B$, computed analytically from the derivatives
of the residual components in
Eqs.~\eqref{eq:18}--\eqref{eq:21}. The zero blocks
$\mathbf{J}_{U\theta}^{(k)} = \mathbf{0}$, $\mathbf{J}_{PP}^{(k)} = 
\mathbf{0}$, $\mathbf{J}_{P\theta}^{(k)} = \mathbf{0}$, and 
$\mathbf{J}_{\theta P}^{(k)} = \mathbf{0}$ reflect, respectively, the 
absence of temperature coupling in the horizontal momentum equation, the 
absence of pressure in the continuity Eq.~\eqref{eq:20}, the absence of 
temperature in the continuity Eq.~\eqref{eq:20}, and the decoupling of 
pressure from the energy Eq.~\eqref{eq:21}. The 
solution is then updated as
\begin{equation}
	\mathbf{U}_h^{(k+1)} = \mathbf{U}_h^{(k)} + 
	\delta\mathbf{U}_h^{(k)},
	\label{eq:25}
\end{equation}
and iterations continue until the convergence criterion 
\begin{equation}
	\left\|\mathbf{R}^{(k)}\right\|_\infty \leq \varepsilon_{\mathbf{R}} 
	= 10^{-10}
	\label{eq:26}
\end{equation}
The linear  systems~\eqref{eq:23} are solved at each iteration 
using a direct LU decomposition.

\subsection{Trust-Region Method}
\label{Sec:3.4}

As an alternative to the standard Newton--Raphson approach, the 
discrete trust-region method
is also employed. In our computational studies, the trust-region 
method achieved better convergence for specific cases involving 
poor initial guesses or ill-conditioned discrete Jacobian matrices, 
particularly at extreme power-law indices $(n \ll 1$ or $n \gg 1)$ 
and high Rayleigh numbers. At each iteration $k$, the algorithm 
constructs a local linear model of the discrete residual
\begin{equation}
	m_k(\boldsymbol{p}) = \mathbf{R}^{(k)} + 
	\mathbf{J}^{(k)}\boldsymbol{p},
	\label{eq:27}
\end{equation}
and approximately minimizes the corresponding trust-region 
least-squares subproblem
\begin{equation}
	\min_{\boldsymbol{p}\,\in\,\mathbb{R}^{N_{\mathrm{dof}}}} 
	\left\|\mathbf{R}^{(k)} + \mathbf{J}^{(k)}\boldsymbol{p}
	\right\|_2^2 \quad \text{subject to} \quad 
	\|\boldsymbol{D}_k\boldsymbol{p}\|_2 \leq \Delta_k,
	\label{eq:28}
\end{equation}
which corresponds to a local quadratic least-squares model for the 
residual norm. Here, $\Delta_k > 0$ is the trust-region radius and 
$\boldsymbol{D}_k$ is a diagonal scaling matrix. In the computations 
reported here, the subproblem is treated by the dogleg strategy used in 
\texttt{NLsolve.jl}: the algorithm first tests the full Gauss--Newton 
step, and if it lies outside the trust region, it uses either the Cauchy 
step or an interpolated step along the dogleg path. The step is accepted 
or rejected based on the ratio of actual to predicted reduction in the 
residual norm
\begin{equation}
	\rho_k = \frac{\left\|\mathbf{R}^{(k)}\right\|_2^2 - 
		\left\|\mathbf{R}^{(k+1)}\right\|_2^2}{\left\|
		\mathbf{R}^{(k)}\right\|_2^2 - \left\|m_k(\boldsymbol{p}_k)
		\right\|_2^2}.
	\label{eq:29}
\end{equation}
If $\rho_k$ is sufficiently positive, the step is accepted and the 
Jacobian is updated at the new iterate; otherwise, the step is rejected 
and the previous iterate is restored. The trust-region radius is reduced 
when the model prediction is poor and enlarged when the prediction is 
accurate and the accepted step is close to the trust-region boundary. 
This adaptive mechanism improves robustness for the strongly 
nonlinear problems arising from the power-law viscosity term 
$\bar{\mu}_{\mathrm{s}}(\mathbf{U}_h)$ in the discrete finite element
formulation, where the standard Newton--Raphson method may fail 
to converge.

\section{Post-processing of heat transfer and flow quantities}\label{Sec:4}

After obtaining the discrete solution $(\boldsymbol{U}_h, P_h, \theta_h)$, key physical quantities are evaluated in a post-processing step. These include the stream function $\psi_h$, the heat function $\Pi_h$, the Nusselt numbers, the entropy generation rates, and the Bejan number, which together characterize the flow structure, the heat transfer, and the thermodynamic irreversibility.

\subsection{Stream Function}
\label{Sec:4.1}

The stream function $\psi_h$ is used to visualize the fluid 
motion through streamlines. For a two-dimensional incompressible 
flow, the velocity components $(U_h, V_h)$ are related to the 
stream function,
\begin{equation}
	U_h = \frac{\partial\psi_h}{\partial Y}, \qquad 
	V_h = -\frac{\partial\psi_h}{\partial X},
	\label{eq:30}
\end{equation}
which, after taking another derivative to the terms in~\eqref{eq:30}, yields the Poisson problem for $\psi_h$, namely, 
\begin{equation}
	\frac{\partial^2\psi_h}{\partial X^2} + 
	\frac{\partial^2\psi_h}{\partial Y^2} = 
	\frac{\partial U_h}{\partial Y} - 
	\frac{\partial V_h}{\partial X}.
	\label{eq:31}
\end{equation}
Therefore, the discrete stream function $\psi_h$ is 
obtained by solving the following weak formulation: Find $\psi_h \in H_0^1(\Omega)$ (where $H_0^1(\Omega)$ denotes a function space with zero boundary conditions) such that for all $\xi \in 
H_0^1(\Omega)$
\begin{equation}
	\int_\Omega \nabla\psi_h \cdot \nabla\xi\,\mathrm{d}\Omega 
	= \int_\Omega \left(\frac{\partial V_h}
	{\partial X} - \frac{\partial U_h}{\partial Y}\right)
	\xi\,\mathrm{d}\Omega
	= \int_\Omega (\boldsymbol{R} \nabla\boldsymbol{U}_h)\,
	\xi\,\mathrm{d}\Omega,
	\label{eq:32}
\end{equation}
with the rotation matrix
\begin{equation}
	\boldsymbol{R} = \begin{pmatrix} 0 & -1 \\ 1 & 0 
	\end{pmatrix}.
	\label{eq:33}
\end{equation}
The homogeneous Dirichlet condition $\psi_h = 0$ on $\partial\Omega$ 
follows directly from the no-slip condition $\boldsymbol{u} = 
\boldsymbol{0}$ on $\partial\Omega$. The discrete problem is solved 
using first-order Lagrange finite elements. By convention (\ref{eq:30}), a 
positive value of $\psi_h$ indicates counter-clockwise circulation, 
while a negative value indicates clockwise circulation.

\subsection{Heat Function}
\label{Sec:4.2}

The heat function $\Pi_h$, similar to $\psi_h$, provides a visualization of the total 
heat flow within the enclosure by combining conductive and 
advective heat fluxes. The total heat flux vector is defined as follows,
\begin{equation}
	\boldsymbol{q}_{\mathrm{tot}} = \boldsymbol{q}_{\mathrm{adv}} 
	+ \boldsymbol{q}_{\mathrm{cond}} = \boldsymbol{U}_h\theta_h 
	- \nabla\theta_h,
	\label{eq:34}
\end{equation}
and the isolines of $\Pi_h$ coincide with the trajectories of 
$\boldsymbol{q}_{\mathrm{tot}}$. Equivalently, the gradient of the heat 
function is orthogonal to the total heat flux.
\begin{equation}
	\nabla\Pi_h \cdot \boldsymbol{q}_{\mathrm{tot}} = 0, 
	\qquad 
	\nabla\Pi_h = \sigma\,\boldsymbol{R}\,
	\boldsymbol{q}_{\mathrm{tot}},
	\label{eq:35}
\end{equation}
with $\boldsymbol{R}$ from Eq.~\eqref{eq:33} and 
$\sigma \in \{-1, +1\}$. To fix the sign convention such that 
$\Pi_h > 0$ for anti-clockwise heat flow~\cite{basakRoleBejansHeatlines2008,basakHeatFlowAnalysis2009,costaBejansHeatlinesMasslines2006}, we set $\sigma = -1$. Taking the negative 
divergence of Eq.~\eqref{eq:35} yields the 
following Poisson problem
\begin{equation}
	-\nabla^2\Pi_h = \nabla\cdot(\boldsymbol{R}\,
	\boldsymbol{q}_{\mathrm{tot}}) = \nabla\cdot\left[
	\boldsymbol{R}(\boldsymbol{U}_h\theta_h - 
	\nabla\theta_h)\right],
	\label{eq:36}
\end{equation}
whose weak formulation reads: given $\theta_h$ and 
$\boldsymbol{U}_h$, find $\Pi_h \in H^1_{\Gamma_{\mathrm{D}}}
(\Omega)$ such that for all $\xi \in H^1_0(\Omega)$
\begin{equation}
	\int_\Omega \nabla\Pi_h \cdot \nabla\xi\,\mathrm{d}\Omega 
	= -\int_\Omega \boldsymbol{R}\left(\boldsymbol{U}_h\theta_h - 
	\nabla\theta_h\right)\cdot\nabla\xi\,
	\mathrm{d}\Omega.
	\label{eq:37}
\end{equation}
The heat function is determined only up to an additive constant unless a 
reference value is prescribed. In the implementation, zero Dirichlet 
values are imposed on the selected heat-function reference tags: these 
are the adiabatic-wall tags for the square-cavity cases and designated 
reference-point tags for the annular cases. Homogeneous natural 
conditions are used on the remaining boundary. This fixes the constant 
and gives a unique discrete heat function. By convention, a positive $\Pi_h$ denotes 
anti-clockwise heat flow and a negative $\Pi_h$ denotes 
clockwise heat flow.

\subsection{Nusselt Number}
\label{Sec:4.3}

The Nusselt number quantifies the dimensionless heat flux at the 
domain boundary. On a boundary $\partial\Omega$ with unit normal 
$\boldsymbol{n}$, the local Nusselt number is defined from the normal 
temperature gradient as
\begin{equation}
	\mathrm{Nu}(\boldsymbol{x}) =
	\left|\frac{\partial \theta_h}{\partial \boldsymbol{n}}\right|
	=
	\left|\nabla\theta_h(\boldsymbol{x})\cdot\boldsymbol{n}\right|.
	\label{eq:38}
\end{equation}
For the square cavity, the local Nusselt numbers on the bottom and 
side walls are therefore
\begin{equation}
	\mathrm{Nu}_b(X) =
	\left|\frac{\partial\theta_h}{\partial Y}(X,0)\right|, 
	\qquad
	\mathrm{Nu}_s(Y) =
	\left|\frac{\partial\theta_h}{\partial X}(X_s,Y)\right|,
	\quad X_s\in\{0,1\}.
	\label{eq:39}
\end{equation}
The corresponding average Nusselt numbers are obtained by integration 
along the respective walls,
\begin{equation}
	\overline{\mathrm{Nu}}_b =
	\int_0^1 \mathrm{Nu}_b(X)\,\mathrm{d}X, \qquad
	\overline{\mathrm{Nu}}_s =
	\int_0^1 \mathrm{Nu}_s(Y)\,\mathrm{d}Y.
	\label{eq:40}
\end{equation}
For the concentric cylindrical annulus, the local and average Nusselt numbers on a 
cylindrical wall is computed from the radial temperature gradient,
\begin{equation}
	\mathrm{Nu}_r(r,\varphi) =
	\left|\frac{\partial\theta_h}{\partial r}(r,\varphi)\right|,
	\quad
	\overline{\mathrm{Nu}}^{\mathrm{in}} =
	\frac{1}{2\pi}\int_0^{2\pi}
	\mathrm{Nu}_r(R_i,\varphi)\,\mathrm{d}\varphi.
	\label{eq:41}
\end{equation}
%

\subsection{Entropy Generation}
\label{Sec:4.4}

Thermodynamic irreversibilities in natural convection systems 
arise from two sources: heat transfer across finite temperature 
gradients and viscous dissipation due to fluid 
friction~\cite{bejan2013entropy}. Following the dimensionless entropy 
generation measures used by Kaluri and Basak~\cite{kaluri2010entropy}, 
the local heat-transfer and fluid-friction contributions in Cartesian 
coordinates $(X, Y)$ are evaluated as
\begin{equation}
	S_{\theta,l} = \left(\frac{\partial\theta_h}{\partial X}
	\right)^2 + \left(\frac{\partial\theta_h}{\partial Y}
	\right)^2,
	\label{eq:43}
\end{equation}
\begin{equation}
	S_{\psi,l} = \phi\left\{2\left[\left(\frac{\partial U_h}
	{\partial X}\right)^2 + \left(\frac{\partial V_h}
	{\partial Y}\right)^2\right] + \left(\frac{\partial U_h}
	{\partial Y} + \frac{\partial V_h}{\partial X}
	\right)^2\right\},
	\label{eq:44}
\end{equation}
where $S_{\theta,l}$ and $S_{\psi,l}$ denote the local entropy 
generation measures due to heat transfer and fluid friction, respectively. 
The parameter $\phi$ is the irreversibility distribution ratio. In its 
dimensional scaling form, it can be written using a reference viscosity 
$\mu_{\mathrm{ref}}$ as
\begin{equation}
	\phi = \frac{\mu_{\mathrm{ref}}\, T_0}{\kappa}\left(\frac{U_0}{\Delta T}
	\right)^2,
	\label{eq:45}
\end{equation}
where $T_0 = (T_h + T_c)/2$ is the bulk reference temperature, 
$U_0 = \alpha/L$ is the reference velocity scale, $\Delta T = 
T_h - T_c$ is the reference temperature difference, and $\kappa$ 
is the thermal conductivity. Following~\cite{kaluri2010entropy}, 
$\phi$ is prescribed as a constant value of $10^{-4}$ in this study, 
rather than recomputed from the local apparent viscosity. This keeps the 
entropy-generation post-processing consistent with the benchmark 
definition and isolates the effect of rheology through the computed 
velocity gradients.

The total entropy generation $S_{\mathrm{total}}$ is obtained 
by summing the contributions from heat transfer 
$(S_{\theta,\mathrm{total}})$ and fluid friction 
$(S_{\psi,\mathrm{total}})$, each evaluated by integrating the 
respective local rates over the domain $\Omega$
\begin{equation}
	S_{\mathrm{total}} = \int_\Omega S_{\theta,l}\,\mathrm{d}\Omega 
	+ \int_\Omega S_{\psi,l}\,\mathrm{d}\Omega = 
	S_{\theta,\mathrm{total}} + S_{\psi,\mathrm{total}},
	\label{eq:46}
\end{equation}
where the integrated forms in terms of the discrete FEM fields are
\begin{equation}
	S_{\theta,\mathrm{total}} = \int_\Omega\left\{
	\left[\frac{\partial}{\partial X}\!\left(\sum_{k=1}^{N_\theta}
	\theta_k\Phi_k^\theta\right)\right]^2 + 
	\left[\frac{\partial}{\partial Y}\!\left(\sum_{k=1}^{N_\theta}
	\theta_k\Phi_k^\theta\right)\right]^2\right\}
	\mathrm{d}X\,\mathrm{d}Y,
	\label{eq:47}
\end{equation}
\begin{equation}
	\begin{aligned}
		S_{\psi,\mathrm{total}} = \phi\int_\Omega\Bigg\{
		2\left[\left(\frac{\partial}{\partial X}\!\sum_{k=1}^{N_u}
		U_k\Phi_k^u\right)^2 + \left(\frac{\partial}{\partial Y}\!
		\sum_{k=1}^{N_u} V_k\Phi_k^u\right)^2\right]
		+ \left(\frac{\partial}{\partial Y}\!\sum_{k=1}^{N_u}
		U_k\Phi_k^u + \frac{\partial}{\partial X}\!\sum_{k=1}^{N_u}
		V_k\Phi_k^u\right)^2&\Bigg\}\mathrm{d}X\,\mathrm{d}Y.
	\end{aligned}
	\label{eq:48}
\end{equation}
All integrals in Eqs.~\eqref{eq:46}--\eqref{eq:48} are evaluated numerically using the numerical framework described in Appendix~\ref{Appendix:B}.

The relative dominance of heat transfer irreversibility over 
viscous dissipation is quantified by the average Bejan number 
$\mathrm{Be}_{\mathrm{av}}$, defined as
\begin{equation}
	\mathrm{Be}_{\mathrm{av}} = \frac{S_{\theta,\mathrm{total}}}
	{S_{\theta,\mathrm{total}} + S_{\psi,\mathrm{total}}} = 
	\frac{S_{\theta,\mathrm{total}}}{S_{\mathrm{total}}}.
	\label{eq:49}
\end{equation}
A value of $\mathrm{Be}_{\mathrm{av}} > 0.5$ indicates that heat transfer 
irreversibility dominates, while $\mathrm{Be}_{\mathrm{av}} < 0.5$ 
indicates dominance of fluid friction irreversibility.

\section{Benchmark Cases}\label{Sec:5}

Prior to presenting the main results, the accuracy and reliability 
of the developed FEM solver are verified through a series of 
benchmark comparisons covering both Newtonian and non-Newtonian 
natural convection flows in square and cylindrical geometries, as well 
as entropy generation predictions. In all validation cases, the 
dimensionless governing Eqs.~\eqref{eq:10}--\eqref{eq:13} are solved subject 
to the corresponding benchmark boundary conditions, and the computed 
quantities are compared against established benchmark solutions from the literature.

\begin{figure} 
	\centering
	\includegraphics[width=1.0\textwidth]{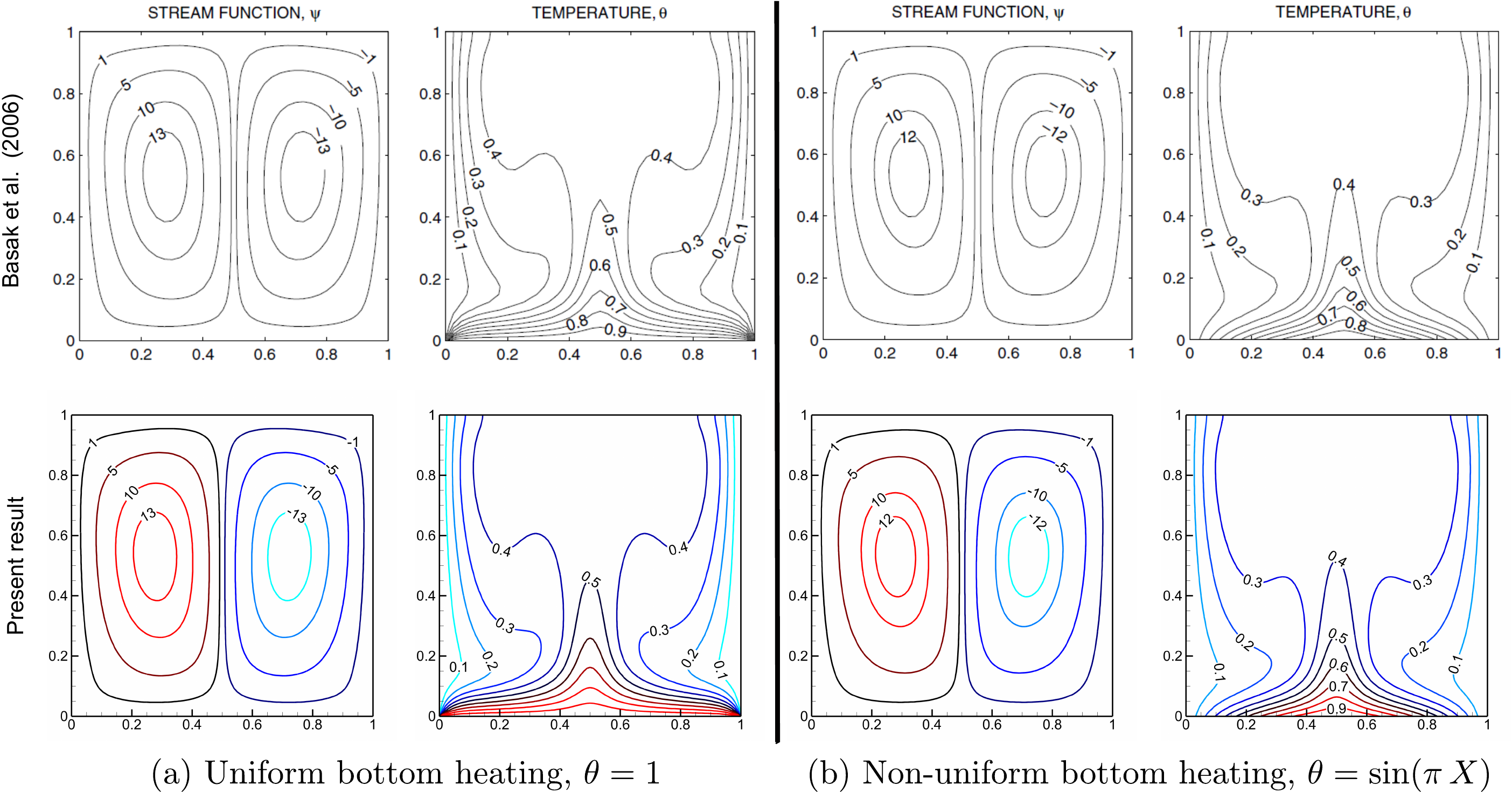}
	\caption{Validation of Newtonian natural convection ($n=1$) in a square
		cavity with bottom-wall heating: (a) uniform heating \((\theta = 1)\) and
		(b) non-uniform sinusoidal heating \((\theta = \sin(\pi X))\). Stream function
		and temperature contours from the present FEM simulations (bottom) are
		compared with the benchmark results of Basak et al.~\cite{basak2006effects}
		(top) for $\operatorname{Pr}=0.7$, and $\operatorname{Ra} = 10^{5}$, demonstrating a good	agreement in both flow and thermal fields.}
	\label{Fig:2}
\end{figure}

\subsection{Newtonian Natural Convection Flows}
\label{Sec:5.1}

As the first benchmark case, Newtonian natural convection ($n=1$) in a square cavity is examined for two bottom-wall thermal boundary conditions, including uniform and non-uniform sinusoidal heating. In both cases, the vertical side walls are kept at a constant cold temperature and the top wall is adiabatic, and the simulations are carried out at $\operatorname{Pr}=0.7$ and $\operatorname{Ra} = 10^{5}$ using a structured grid of \(200 \times 200\) quadrilateral elements. Figure~\ref{Fig:2} presents the stream function and temperature contours from the present FEM solver alongside the benchmark results of Basak et al.~\cite{basak2006effects}. For uniform heating, the symmetric thermal forcing produces a pair of counter-rotating circulation cells, with pronounced thermal boundary layers developing near the hot bottom and cold side walls, whereas under non-uniform sinusoidal heating the circulation remains symmetric but is driven more strongly by the localized heating at the center of the bottom wall, where the isotherms rise sharply and indicate a strong buoyant plume. A good agreement in both flow patterns and thermal fields for the two heating configurations demonstrates the accuracy and robustness of the present FEM solver.

\begin{figure} 
	\centering
	\includegraphics[width=1.0\textwidth]{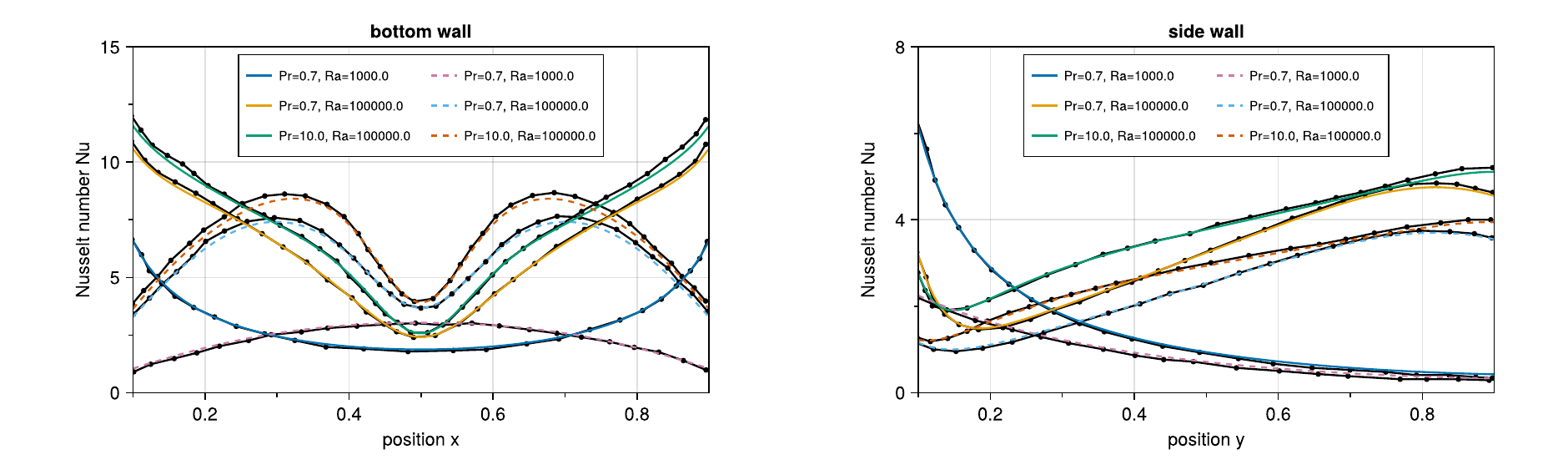}
	\caption{Validation of Newtonian natural convection ($n=1$) in a square cavity with bottom-wall uniform and non-uniform heating. Distributions of the local Nusselt number are shown along (a) the bottom wall and (b) the side wall for $\operatorname{Pr} = 0.7$, $10$ and $\operatorname{Ra} = 10^3$, $10^5$. Solid lines represent uniform heating, while dashed lines correspond to non-uniform sinusoidal heating. Present results are compared with benchmark data of Basak et al.~\cite{basak2006effects}, shown by symbols.}
	\label{Fig:3}
\end{figure}

Figure~\ref{Fig:3} provides a quantitative validation of the local heat transfer by comparing the local Nusselt number distributions along the bottom and side walls of the square cavity with the benchmark
data of Basak et al.~\cite{basak2006effects}. Results are shown for
\(\operatorname{Pr} = 0.7, 10\) and \(\operatorname{Ra} = 10^{3}, 10^{5}\) under both uniform and non-uniform sinusoidal bottom-wall heating.
Along the bottom wall, uniform heating produces a symmetric profile with peak Nusselt numbers near the side walls, whereas non-uniform heating shifts the peak toward the centrally heated region; increasing \(\operatorname{Ra}\)
enhances the gradients and raises the local Nusselt levels. Along the side wall, the highest heat transfer occurs near the bottom corner,  especially at larger \(\operatorname{Ra}\), due to strong thermal plumes from the heated
bottom. The present curves closely match the trends and magnitudes of
Basak et al.~\cite{basak2006effects}, confirming the solver's accuracy for local heat transfer prediction.

\begin{figure}
	\centering
	\includegraphics[width=0.7\textwidth]{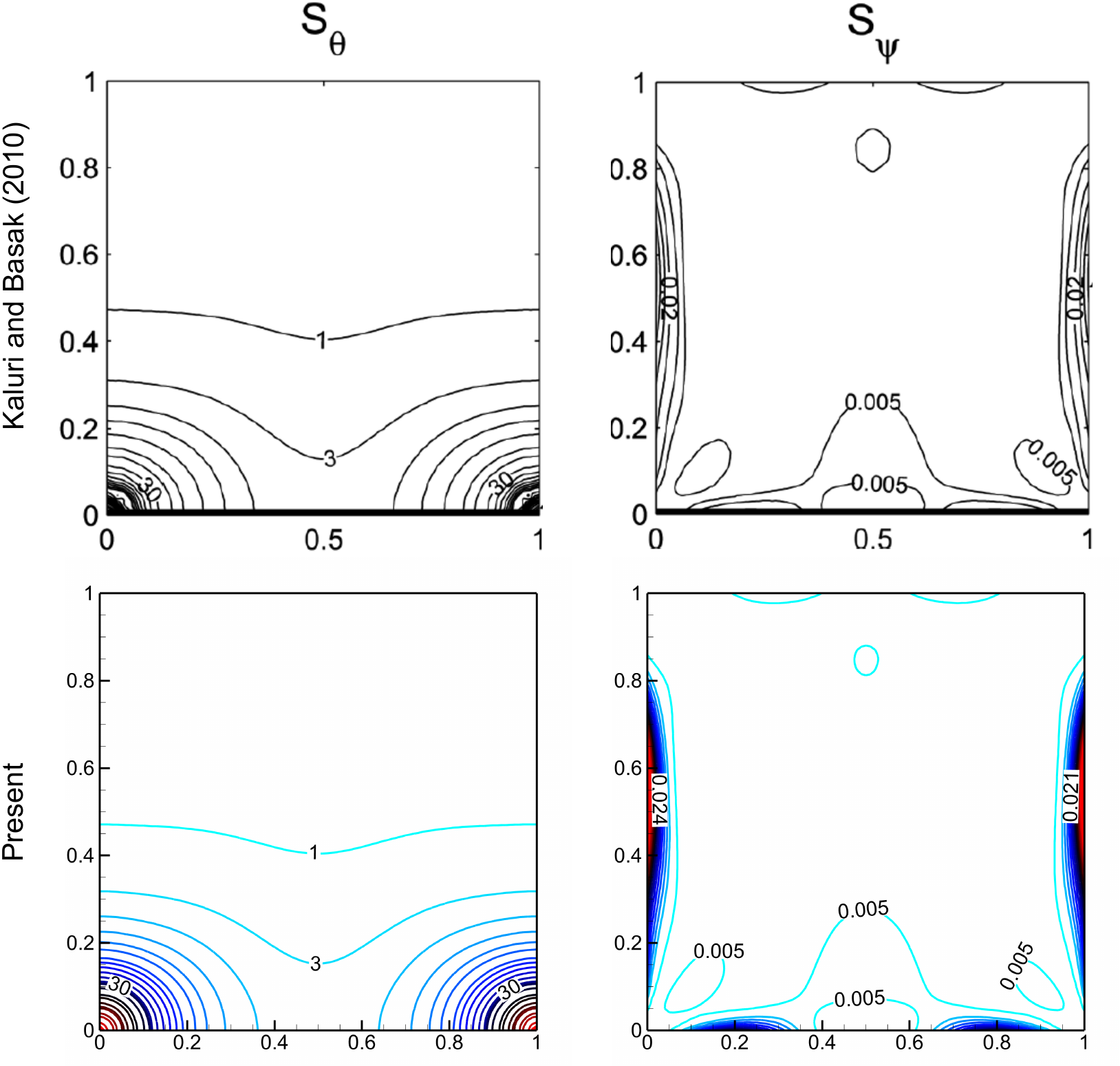}
	\caption{Validation of Newtonian natural convection ($n=1$) in a square cavity with uniform heated bottom wall. Local entropy generation contours due to heat transfer $S_{\theta,l}$ and fluid friction  $S_{\psi,l}$ are shown at $\operatorname{Pr} = 0.015$ and $\operatorname{Ra} = 10^3$. Present FEM results are compared against the numerical data of Kaluri and Basak~\cite{kaluri2010entropy}.}
	\label{Fig:4}
\end{figure}

To validate the entropy-generation post-processing framework, a square cavity with a uniformly heated bottom wall is simulated at \(\operatorname{Pr} = 0.015\) and \(\operatorname{Ra} = 10^{3}\), following the configuration of Kaluri and Basak~\cite{kaluri2010entropy}. Figure~\ref{Fig:4} shows the
local entropy generation contours due to heat transfer \(S_{\theta,l}\) and fluid friction \(S_{\psi,l}\) from the present FEM solver together with the reference results of Kaluri and Basak~\cite{kaluri2010entropy}. Heat-transfer
entropy production is concentrated on the thermally active walls and corner regions where temperature gradients are largest, whereas viscous entropy production is localized near the cavity corners where velocity gradients are highest. The close agreement in both spatial patterns and relative magnitudes of \(S_{\theta,l}\) and \(S_{\psi,l}\) confirms the accuracy of the proposed entropy post-processing framework.

\begin{figure}
	\centering
		\includegraphics[width=0.7\textwidth]{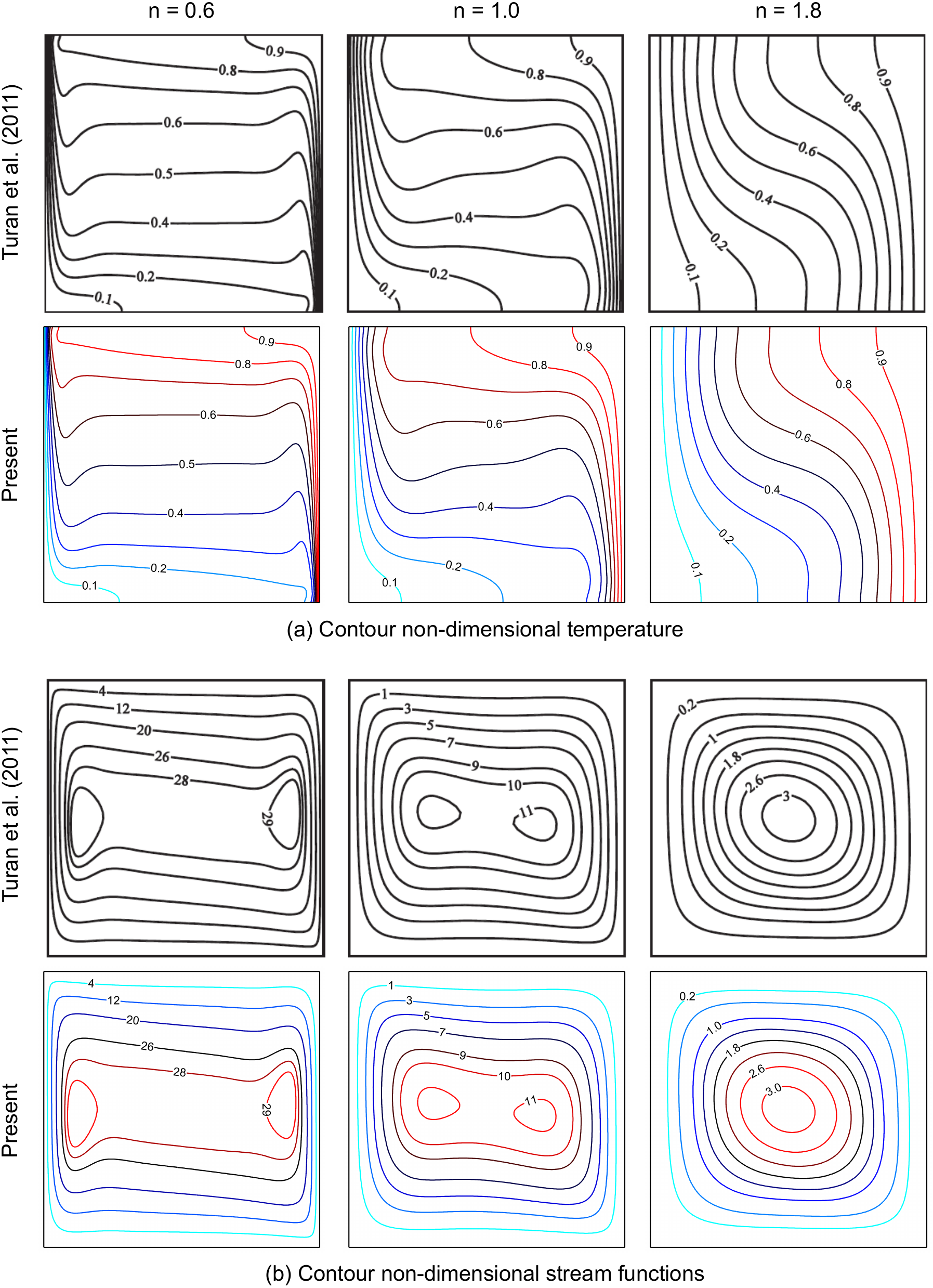}
		\caption{Validation of non-Newtonian natural convection in a differentially heated square cavity for $\operatorname{Pr}=10^3$, $\operatorname{Ra}=10^4$, and power-law indices $n = 0.6, 1.0$, and $1.8$: comparison of non-dimensional (a) temperature and (b) stream function contours between Turan et al.~\cite{turan2011laminar} and the present FEM solver.}
	\label{Fig:5}
\end{figure}

\begin{figure}
	\centering
		\includegraphics[width=1.0\textwidth]{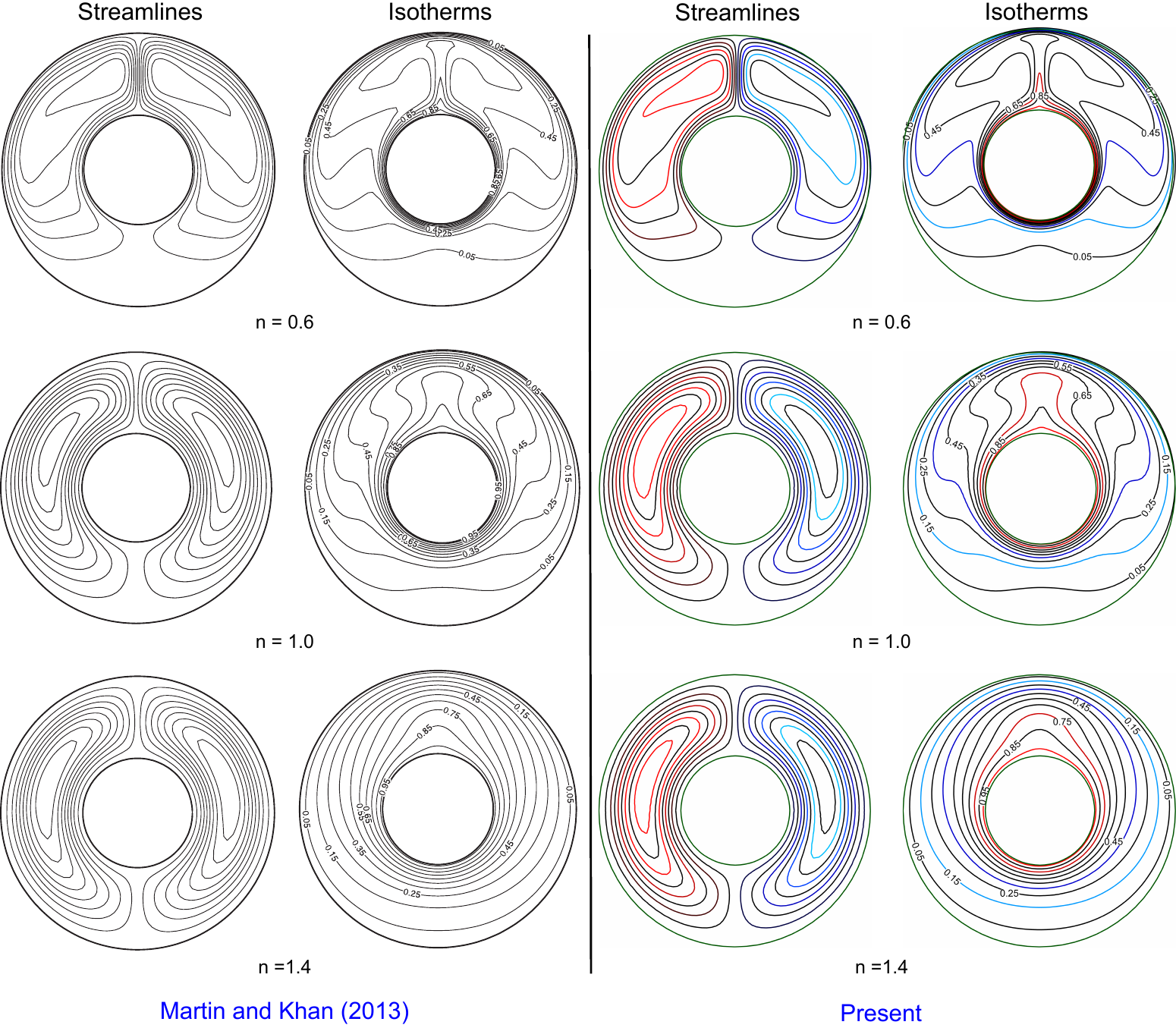}
		\caption{Validation of non-Newtonian natural convection in a concentric cylindrical annulus for $\operatorname{Pr}=100$, $\operatorname{Ra}=10^4$, and power-law indices 
			$n = 0.6, 1.0$, and $n = 1.4$. Comparison of streamline and isotherm contours between Matin and Khan~\cite{matin2013laminar} and the present FEM results.}
	\label{Fig:6}
\end{figure}

\subsection{Non-Newtonian Natural Convection Flows}
\label{Sec:5.2}

The next benchmark examines non-Newtonian natural convection of power-law fluids in a square cavity using the present FEM solver, and is compared with the numerical results of Turan et al.~\cite{turan2011laminar}. In the square cavity, the vertical walls are held at different temperatures, while the horizontal walls are adiabatic. Both velocity components vanish on all boundaries, enforcing no-slip and impermeable conditions. The simulations are conducted at \(\operatorname{Pr}=10^3\) and \(\operatorname{Ra}=10^4\) for three different power-law indices ($n=0.6, 1.0$, and $1.8$). Figure~\ref{Fig:5} shows the corresponding non-dimensional temperature and
stream function contours. For $n=0.6$, the reduced apparent viscosity enhances convection, leading to steep thermal gradients near the vertical walls and stronger circulation. At $n=1.0$, the temperature field is more uniform and a single primary vortex dominates the cavity, as expected for Newtonian flow. For \(n = 1.8\), the increased viscosity damps motion, yielding weaker
circulation and more diffuse, nearly conduction-dominated isotherms. The present FEM predictions closely match those of Turan
et al.~\cite{turan2011laminar} for all \(n\), demonstrating that the solver accurately captures power-law rheology effects on non-Newtonian natural convection.

The final benchmark examines non-Newtonian natural convection in a concentric
cylindrical annulus, using the present FEM solver and the reference solutions of
Matin and Khan~\cite{matin2013laminar}. The outer cylinder is kept at a uniform
cold temperature ($T_{c}$), while the inner cylinder is maintained at a uniform hot temperature ($T_{h}$) with ($T_{c} < T_{h}$). The simulations are performed at \(\operatorname{Pr}=100\) and \(\operatorname{Ra}=10^4\) for three power-law indices, $n=0.6, 1.0$, and $1.4$. 
Figure~\ref{Fig:6} illustrates the corresponding streamline and isotherm contours. For \(n = 0.6\), the reduced apparent viscosity strengthens convection near the heated inner cylinder,
generating compact vortices and strongly distorted isotherms that indicate
vigorous radial heat transport. At \(n = 1.0\), the flow becomes more symmetric
and the isotherms are only moderately displaced from the purely conductive
pattern, typical of Newtonian convection. For \(n = 1.4\), the higher apparent
viscosity substantially damps circulation, and the isotherms become nearly
concentric, signaling conduction-dominated transport. The present FEM
predictions match those of Matin and Khan~\cite{matin2013laminar} for both
velocity and temperature fields across all three rheological regimes,
demonstrating the robustness and accuracy of the proposed framework for
non-Newtonian natural convection in cylindrical geometries.

\section{Numerical results and discussion}\label{Sec:6}

This section presents the numerical results obtained using the FEM solver
for the square cavity and concentric cylindrical annulus configurations.
The effects of thermal boundary conditions and power-law rheology on the
flow structure, heat transfer, and entropy generation are examined
systematically. Particular attention is given to the influence of the
Rayleigh number and the power-law index on the transport characteristics
in both geometries. The simulations are performed at \(\operatorname{Pr} = 100\),
\(\operatorname{Ra} = 10^{4}\), and for \(n = 0.6\), \(1.0\), and \(1.4\). Flow
behavior is characterized using streamlines and heatlines, heat-transfer
performance is evaluated through local and average Nusselt numbers, and
thermodynamic irreversibility is assessed from entropy generation due to
heat transfer and viscous dissipation. 

\subsection{Grid independence study}
\label{Sec:6.1}
\begin{figure}
	\centering
	\includegraphics[width=0.7\textwidth]{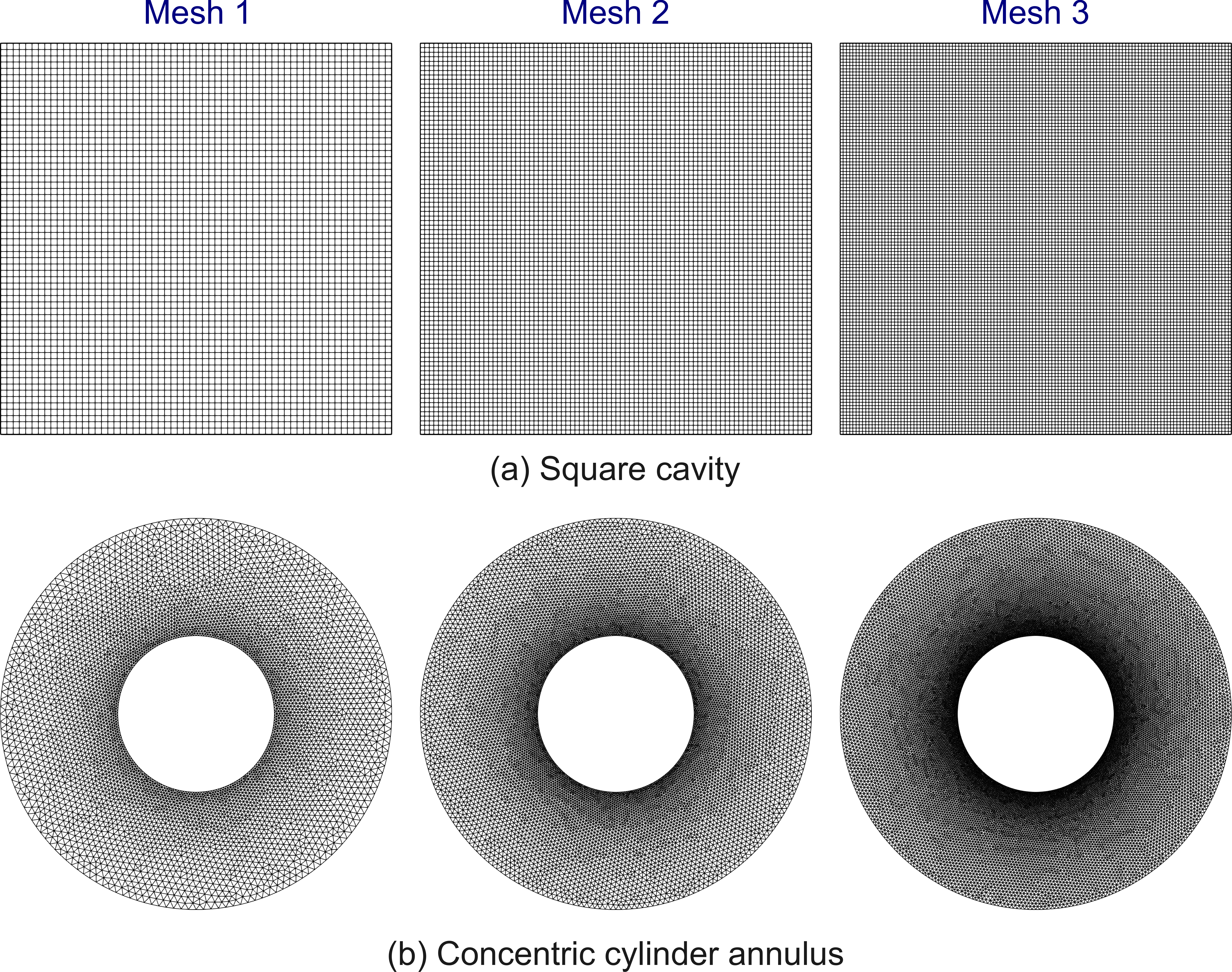}
	\caption{Mesh convergence study for the two computational domains: (a) structured quadrilateral meshes (Mesh 1--3) for the square cavity, and (b) unstructured triangular meshes (Mesh 1--3) for the concentric cylindrical annulus. Three progressively refined mesh resolutions are compared to assess grid independence and ensure numerical accuracy.}
	\label{Fig:7}
\end{figure}
A mesh-independence study is conducted for both geometries to ensure
grid-independent solutions. Three mesh resolutions are examined, illustrated in Fig.~\ref{Fig:7}: structured quadrilateral meshes are used for the square cavity, whereas unstructured meshes with radial refinement are
employed for the concentric cylindrical annulus. The average Nusselt number at the heated wall, maximum local entropy 
generation due to heat transfer ($S_{\theta,\max}$), and fluid friction 
($S_{\psi,\max}$) are monitored across all mesh levels 
(Tables~\ref{tab:Table_1} and~\ref{tab:Table_2}). For the square cavity, all three quantities exhibit negligible variation beyond 
Mesh 3 (Table~\ref{tab:Table_1}). A consistent trend is observed for the 
concentric cylinder annulus, where $\overline{\text{Nu}}_{\text{inner}}$, 
$S_{\theta,\max}$, and $S_{\psi,\max}$ stabilize from Mesh 3 onward 
(Table~\ref{tab:Table_2}). Accordingly, Mesh 3 is adopted for all subsequent 
simulations, providing an optimal balance between computational efficiency and 
solution accuracy.

\begin{table}[h!]
	\centering
	\caption{Grid independence study for the square cavity: variation of the average Nusselt number at the bottom wall ($\overline{\text{Nu}}_{\text{bot}}$), maximum 
		local entropy generation due to heat transfer ($S_{\theta,\text{max}}$), and 
		fluid friction ($S_{\psi,\text{max}}$) with mesh refinement.}
	\label{tab:Table_1}
	\begin{tabular}{cccccc}
		\textbf{id} & $n_{\text{elems}}$ & $n_{\text{nodes}}$ & 
		$\overline{\text{Nu}}_{\text{bot}}$ & $S_{\theta,\text{max}}$ & $S_{\psi,\text{max}}$ \\
		\hline
		1 & 3844  & 3969  & 6.6704 & 104.3720 & 2406.6836 \\
		2 & 7396  & 7569  & 6.6460 & 103.5721 & 2450.5564 \\
		3 & 14884 & 15129 & 6.6312 & 103.0823 & 2475.2736 \\
		4 & 29584 & 29929 & 6.6232 & 102.8219 & 2487.4909 \\
		5 & 59536 & 60025 & 6.6189 & 102.6823 & 2493.9131 \\
		\hline
	\end{tabular}
\end{table}

\begin{table}[h!]
	\centering
	\caption{Grid independence study for the concentric cylinder annulus: variation of the average Nusselt number at the inner wall ($\overline{\text{Nu}}_{\text{inner}}$), maximum local entropy generation due to heat transfer ($S_{\theta,\text{max}}$), and fluid friction ($S_{\psi,\text{max}}$) with mesh refinement.}
	\label{tab:Table_2}
	\begin{tabular}{cccccc}
		\hline
		ID & $n_{\text{elems}}$ & $n_{\text{nodes}}$ & $\text{Nu}_{\text{inner\_avg}}$ & $S_{\theta,\max}$ & $S_{\psi,\max}$ \\
		\hline
		1 & 10254  & 5306   & 4.0284  & 55.6149  & 1988.3033 \\
		2 & 19976  & 10240  & 4.0224  & 56.0836  & 1980.8299 \\
		3 & 40012  & 20363  & 4.0215  & 56.2028  & 2035.7779 \\
		4 & 79434  & 40221  & 4.0221  & 56.1087  & 1726.6153 \\
		5 & 158098 & 79761  & 4.0218  & 56.1371  & 1720.5476 \\
		\hline
	\end{tabular}
\end{table}

\begin{figure}
	\centering
	\includegraphics[width=0.8\textwidth]{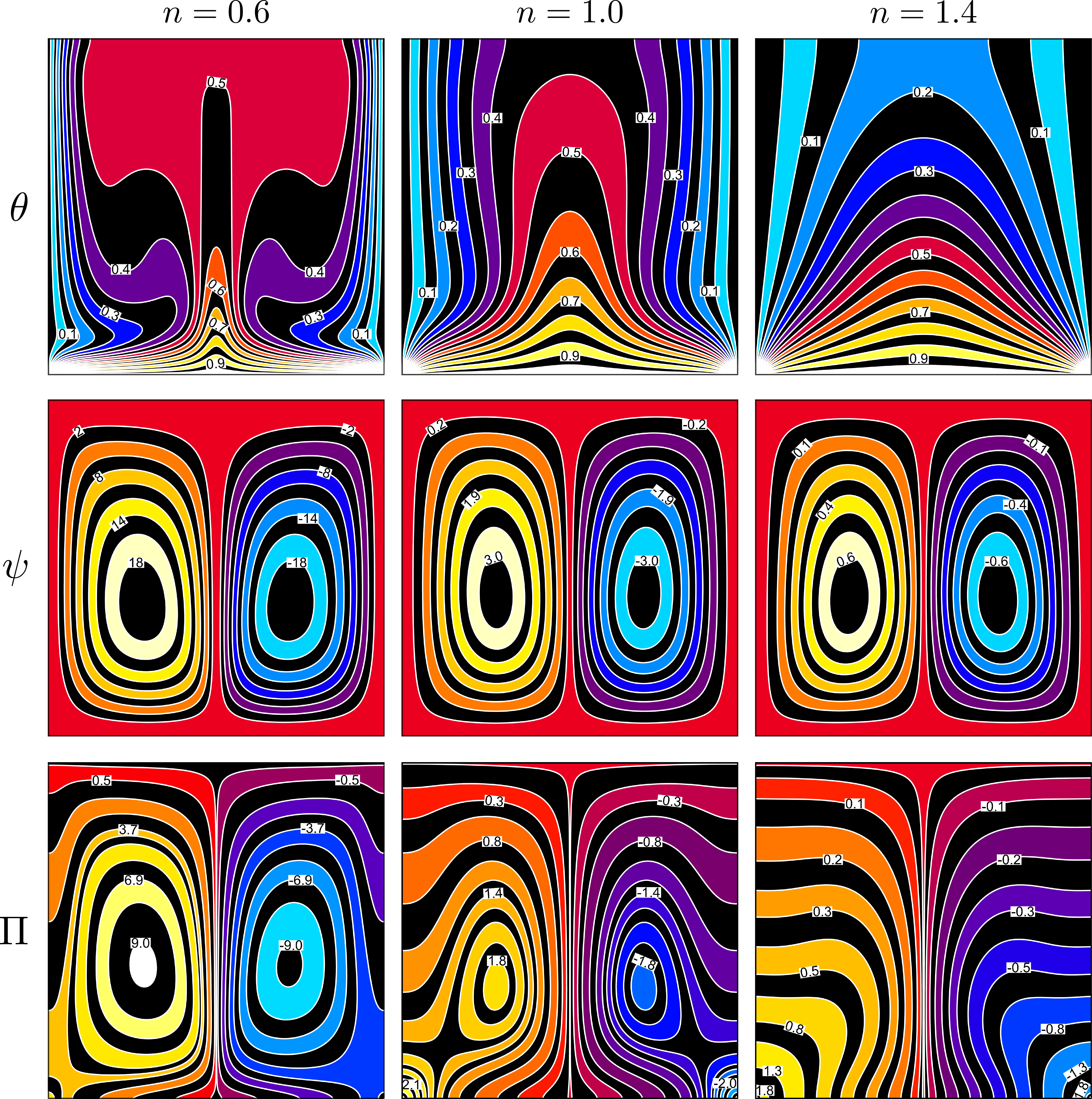}
	\caption{Isotherms $\theta$ (top), streamlines $\psi$ (middle), and heatlines $\Pi$ (bottom) of power-law index $n = 0.6$, $1.0$, and $1.4$ in a square cavity with uniform bottom-wall heating ($\theta = 1$).}
	\label{Fig:8}
\end{figure}

\subsection{Square cavity configuration}
\label{Sec:6.2}

This subsection examines the effects of the power-law index $n$ and thermal 
boundary conditions on non-Newtonian natural convection in the square cavity, considering both uniform and non-uniform sinusoidal bottom-wall heating at $\operatorname{Pr} = 100$ and $\operatorname{Ra} = 10^4$.

\subsubsection{Flow structure and thermal fields}
\label{Sec:6.2.1}

Figure~\ref{Fig:8} illustrates the isotherms $\theta$, streamlines $\psi$, 
and heatlines $\Pi$ for $n = 0.6$, $1.0$, and $1.4$ in the square cavity 
with uniform bottom-wall heating ($\theta = 1$). For the 
shear-thinning fluid ($n = 0.6$), the reduced apparent viscosity enhances 
buoyancy-driven circulation, producing two strong counter-rotating vortices 
that occupy the lower portion of the cavity, while the isotherms exhibit 
steep thermal gradients near the heated bottom wall and a thermally 
stratified core, indicating vigorous convective transport. At $n = 1.0$, 
the flow transitions to a more symmetric pattern with expanded vortex 
structures and moderately distorted isotherms, consistent with classical 
Newtonian natural convection. For the shear-thickening fluid ($n = 1.4$), 
the increased apparent viscosity suppresses fluid motion considerably, 
yielding weaker and more diffuse streamlines and nearly parallel isotherms 
that reflect a conduction-dominated thermal field. The heatlines confirm 
these trends: concentrated and curved $\Pi$-contours for $n = 0.6$ indicate 
strong convective heat flux from the bottom wall, which progressively 
weakens and becomes more uniform as $n$ increases toward shear-thickening 
behavior.

\begin{figure}
	\centering
	\includegraphics[width=0.8\textwidth]{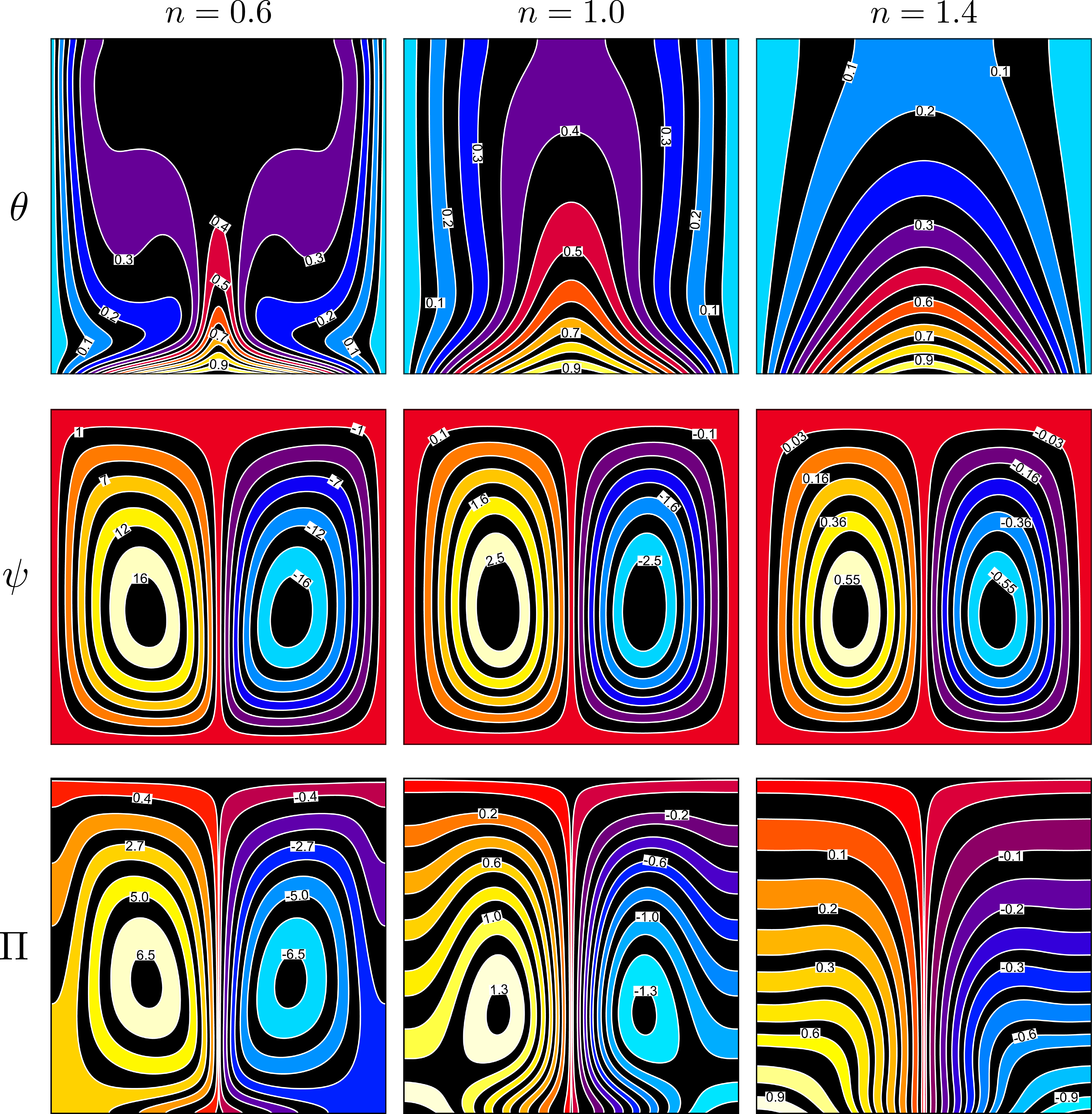}
	\caption{Isotherms $\theta$ (top), streamlines $\psi$ (middle), and heatlines $\Pi$ (bottom) of power-law index $n = 0.6$, $1.0$, and $1.4$ in a square cavity with non-uniform bottom-wall heating ($\theta=\sin (\pi \, X)$).}
	\label{Fig:9}
\end{figure}

Figure~\ref{Fig:9} presents the isotherms $\theta$, streamlines $\psi$, and 
heatlines $\Pi$ for $n = 0.6$, $1.0$, and $1.4$ under non-uniform sinusoidal 
bottom-wall heating ($\theta = \sin(\pi X)$). 
Compared to the uniform heating case (Fig.~\ref{Fig:8}), the sinusoidal 
boundary condition introduces a spatially localized thermal forcing at the 
cavity center, which breaks the symmetric thermal stratification and generates 
a distinct central buoyant plume visible in the isotherms for all values of 
$n$. For $n = 0.6$, the shear-thinning effect amplifies this plume-driven 
convection, producing tightly packed streamlines and strongly distorted 
isotherms concentrated near the bottom center, whereas the heatlines indicate 
intense convective heat flux directed upward from the peak-heated region. At 
$n = 1.0$, the flow retains two counter-rotating vortices but with reduced 
intensity compared to the shear-thinning case, and the isotherms show moderate 
thermal gradients consistent with Newtonian behavior under localized heating. 
For $n = 1.4$, viscous damping significantly weakens circulation, and the 
isotherms approach a near-conductive distribution; however, unlike the uniform 
heating case, a residual central plume persists due to the concentrated 
sinusoidal heat input, and the heatlines become increasingly diffuse, 
reflecting the dominance of conduction over convection in the shear-thickening 
regime.

\begin{figure} 
	\centering
	\includegraphics[width=0.8\textwidth]{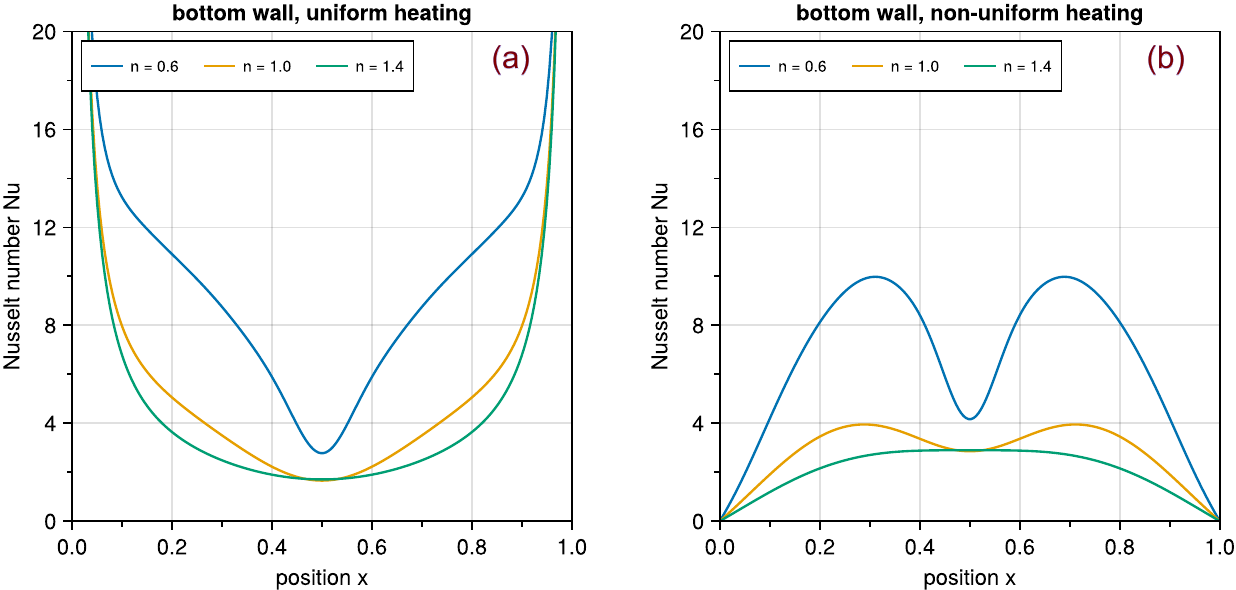}
	\caption{Local Nusselt number distributions along the bottom wall of the 
		square cavity for power-law indices $n = 0.6$, $1.0$, and $1.4$ under 
		(a) uniform heating ($\theta = 1$) and (b) non-uniform sinusoidal heating ($\theta = \sin(\pi X)$).}
	\label{Fig:10}
\end{figure}

\subsubsection{Nusselt number and heat transfer}
\label{Sec:6.2.2}

Figure~\ref{Fig:10} compares the local Nusselt number distributions along 
the bottom wall for $n = 0.6$, $1.0$, and $1.4$ under uniform and 
non-uniform sinusoidal heating. Under 
uniform heating, the local $Nu$ exhibits a symmetric U-shaped profile, with 
peak values concentrated near the side walls due to intense thermal boundary 
layer development driven by the cold vertical walls, and a minimum at the 
cavity center where convective mixing is relatively weak. In 
contrast, non-uniform sinusoidal heating ($\theta = \sin(\pi X)$) reverses 
this trend, producing an M-shaped profile with two local maxima near $X = 
0.25$ and $X = 0.75$ and a reduced value at the center, consistent with the 
localized buoyant plume generated by the sinusoidal temperature peak at 
$X = 0.5$. In both cases, the shear-thinning fluid ($n = 
0.6$) yields the highest $Nu$ values owing to its reduced apparent viscosity, 
which enhances convective transport, while the shear-thickening fluid ($n = 
1.4$) produces the lowest heat transfer rates across the entire wall due to 
viscous suppression of fluid motion. Notably, the difference in 
$Nu$ between rheological regimes is more pronounced under non-uniform heating, 
suggesting that sinusoidal thermal forcing amplifies the sensitivity of heat 
transfer to power-law rheology compared to the uniform heating case.

\begin{figure} 
	\centering
	\includegraphics[width=0.8\textwidth]{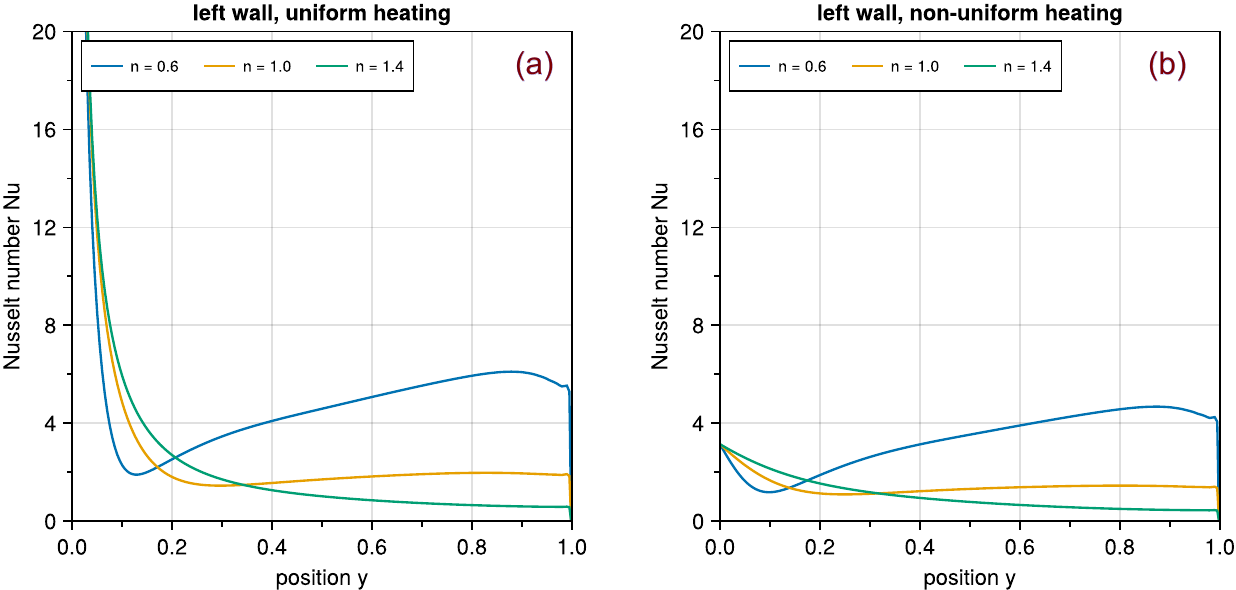}
	\caption{Local Nusselt number distributions along the left wall of the 
		square cavity for power-law indices $n = 0.6$, $1.0$, and $1.4$ under 
		(a) uniform heating ($\theta = 1$) and (b) non-uniform sinusoidal heating ($\theta = \sin(\pi X)$).}
	\label{Fig:11}
\end{figure}

Figure~\ref{Fig:11} presents the local Nusselt number distributions along 
the left (cold) side wall of the square cavity for $n = 0.6$, $1.0$, and 
$1.4$ under uniform and non-uniform sinusoidal bottom-wall heating ($\theta = \sin(\pi X)$). Under uniform heating, the local $Nu$ along 
the side wall exhibits its maximum near the bottom corner, where the rising 
thermal plume from the uniformly heated bottom wall impinges directly on the 
cold vertical surface, generating steep thermal gradients; the heat transfer 
rate decays monotonically toward the adiabatic top wall as the thermal 
boundary layer thickens with increasing height. Under non-uniform sinusoidal 
heating, the side wall $Nu$ distribution retains a similar decreasing trend 
from bottom to top but with a notably lower peak value compared to the 
uniform case, since the sinusoidal heating concentrates thermal energy at 
the cavity center rather than uniformly energizing the bottom wall, thereby 
reducing the thermal gradient at the bottom corner of the side wall. Across 
both heating conditions, the shear-thinning fluid ($n = 0.6$) produces the 
highest local \textit{Nu} values along the side wall, driven by its enhanced 
convective circulation, while the shear-thickening fluid ($n = 1.4$) yields 
the lowest heat transfer rates due to viscous suppression; however, the 
rheological spread between the three $n$-values is more pronounced under 
uniform heating, confirming that uniform thermal loading engages the full 
convective capacity of the cavity more effectively than localized sinusoidal 
forcing.

\begin{figure}
	\centering
	\includegraphics[width=0.8\textwidth]{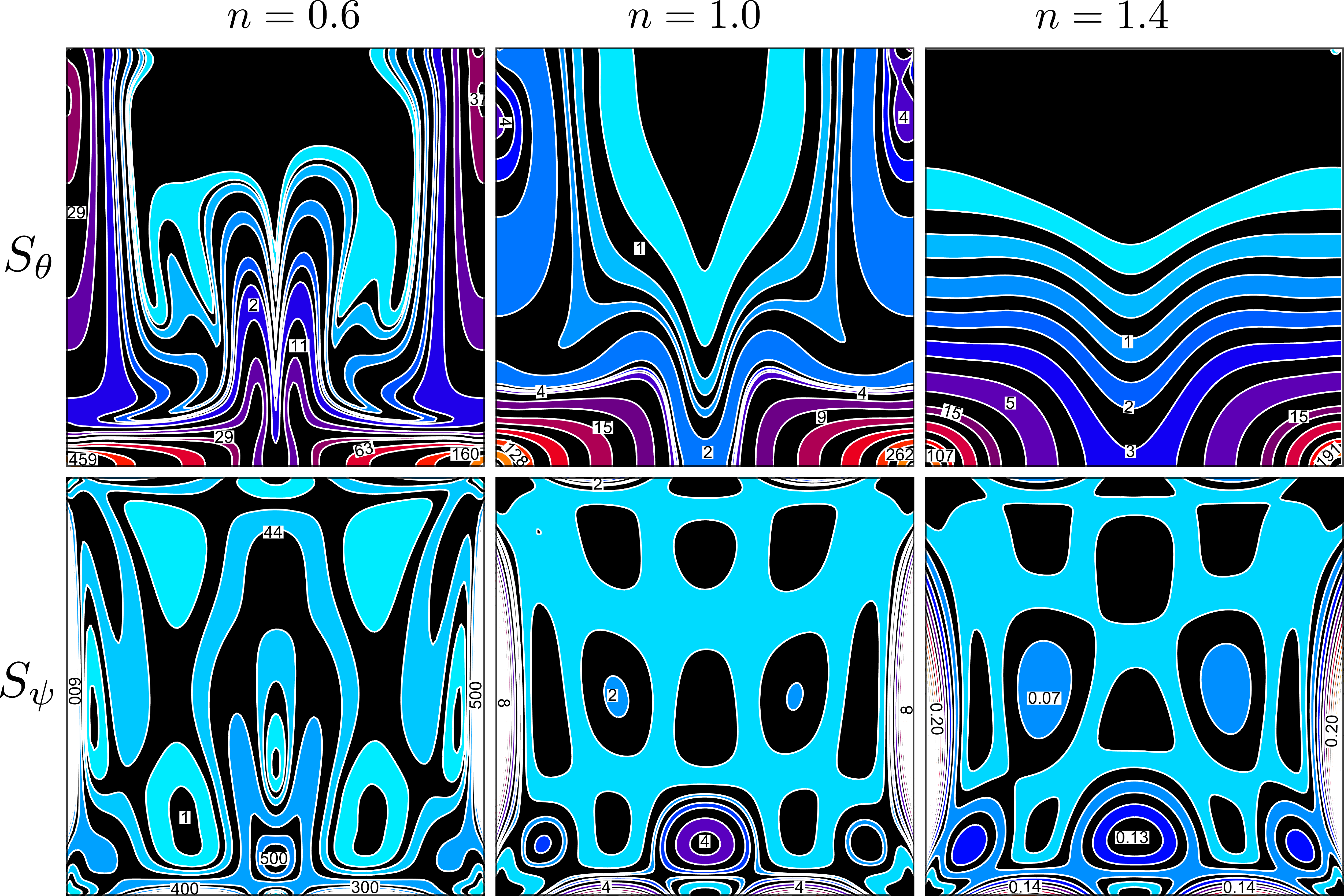}
	\caption{Contours of local entropy generation due to heat transfer 
		$S_{\theta}$ (top) and fluid friction $S_{\psi}$ (bottom) for 
		power-law indices $n = 0.6$, $1.0$, and $1.4$ in the square cavity 
		with uniform bottom-wall heating ($\theta = 1$).}
	\label{Fig:12}
\end{figure}

\subsubsection{Entropy generation and Bejan number}
\label{Sec:6.2.3}

Figure~\ref{Fig:12} presents the contours of local entropy generation due 
to heat transfer $S_{\theta}$ and fluid friction $S_{\psi}$ for $n = 
0.6$, $1.0$, and $1.4$ under uniform bottom-wall heating. The $S_{\theta}$ contours are most concentrated near the cold 
vertical side walls and along the heated bottom wall, regions that correspond 
directly to the steep isotherm gradients observed in Fig.~\ref{Fig:8}, where 
the thermal boundary layers are thinnest and the temperature difference 
between fluid and wall is greatest. For $n = 0.6$, the 
shear-thinning fluid exhibits the highest $S_{\theta}$ magnitudes, 
consistent with the strongly distorted isotherms and vigorous counter-rotating 
vortices seen in the streamline contours of Fig.~\ref{Fig:8}, which intensify 
thermal gradients at the walls and thereby amplify heat transfer 
irreversibility. As $n$ increases to $1.4$, the weakened 
circulation evident in the streamlines leads to a more diffuse $S_{\theta}$ 
distribution and reduced peak magnitudes, reflecting the conduction-dominated 
thermal field. The $S_{\psi}$ contours are highly localized near the cavity corners and the bottom wall, where the tightly packed streamlines in Fig.~\ref{Fig:8} indicate peak velocity gradients; these magnitudes decrease progressively with increasing $n$, as shear-thickening rheology suppresses fluid motion and reduces viscous dissipation. Hence, $S_{\theta}$ dominates over $S_{\psi}$ across all rheological regimes, confirming that heat transfer irreversibility is the primary source of thermodynamic inefficiency in buoyancy-driven power-law fluid convection at the considered $\operatorname{Ra}$.

\begin{figure}
	\centering
	\includegraphics[width=0.8\textwidth]{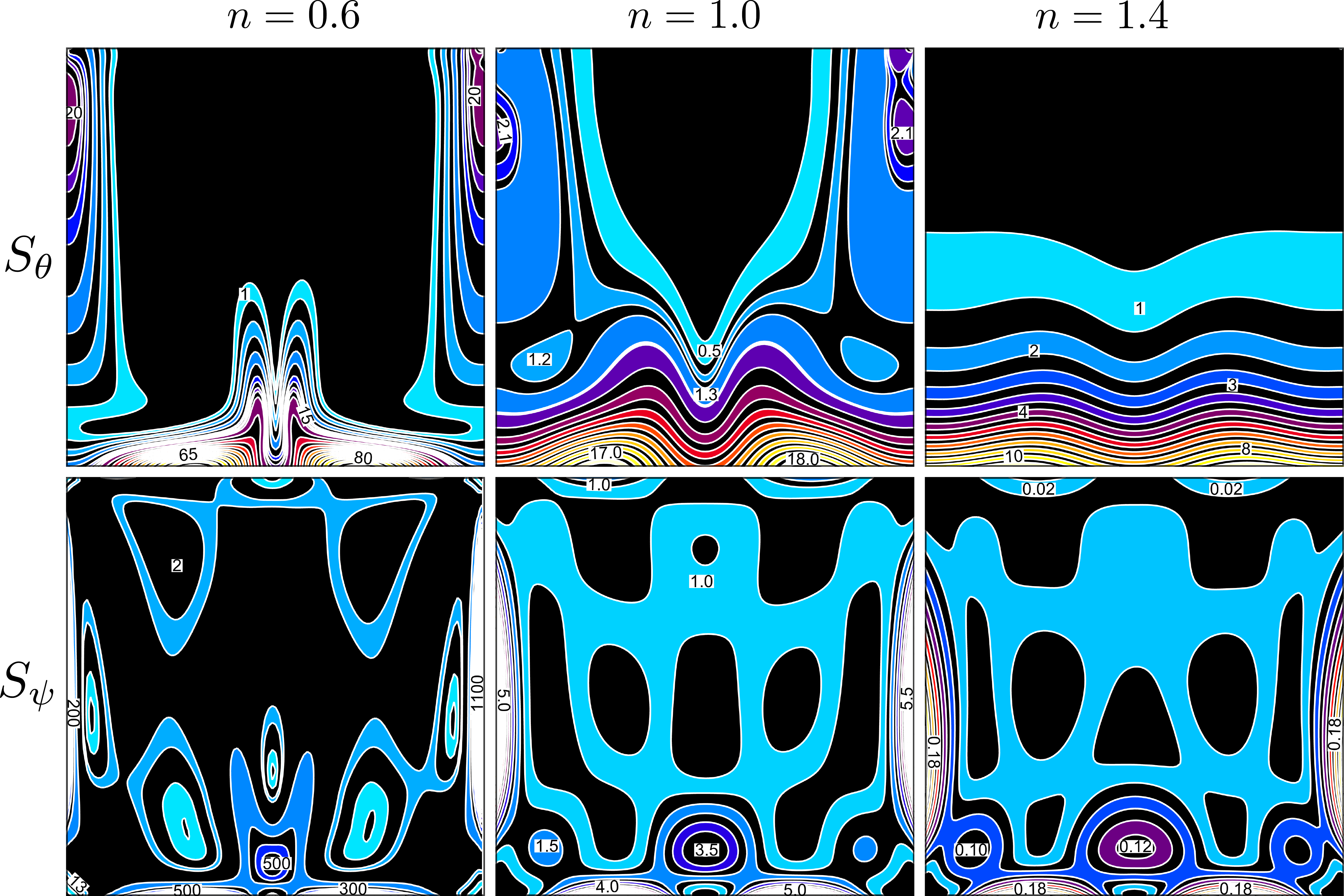}
	\caption{Contours of local entropy generation due to heat transfer 
		$S_{\theta}$ (top) and fluid friction $S_{\psi}$ (bottom) for 
		power-law indices $n = 0.6$, $1.0$, and $1.4$ in the square cavity 
		with non-uniform bottom-wall heating ($\theta = \sin(\pi X)$).}
	\label{Fig:13}
\end{figure}

Figure~\ref{Fig:13} presents the contours of local entropy generation due 
to heat transfer $S_{\theta}$ and fluid friction $S_{\psi}$ for $n = 
0.6$, $1.0$, and $1.4$ under non-uniform sinusoidal bottom-wall heating 
($\theta = \sin(\pi X)$). In contrast to the 
uniform heating case (Fig.~\ref{Fig:12}), the $S_{\theta}$ contours are no 
longer uniformly distributed along the bottom wall but instead exhibit a 
spatially concentrated pattern directly above the central peak of the 
sinusoidal temperature profile, consistent with the localized buoyant plume 
and sharply rising isotherms observed at the cavity center in 
Fig.~\ref{Fig:9}. For $n = 0.6$, the reduced apparent viscosity intensifies 
the central plume-driven convection, producing the highest $S_{\theta}$ 
magnitudes among all rheological cases, with peak irreversibility localized 
in the plume column where thermal gradients are steepest; this is 
corroborated by the tightly clustered isotherms and strong upward convective 
motion seen in the streamlines of Fig.~\ref{Fig:9}. As $n$ increases to 
$1.4$, the weakened circulation suppresses the central plume, leading to a 
considerably more diffuse $S_{\theta}$ field with significantly reduced peak 
values compared to both the shear-thinning case and the corresponding uniform 
heating scenario in Fig.~\ref{Fig:12}, reflecting the dominance of conduction 
in the shear-thickening regime. The $S_{\psi}$ contours remain localized near 
the cavity corners and bottom wall edges across all $n$-values, where the 
streamlines in Fig.~\ref{Fig:9} indicate the highest velocity gradients; 
however, the $S_{\psi}$ magnitudes under non-uniform heating are notably 
lower than those in the uniform heating case, since the sinusoidal forcing 
concentrates fluid motion near the cavity center rather than energizing the 
full bottom wall, thereby reducing overall fluid friction. 

\begin{table}[h!]
	\centering
	\caption{Total entropy generation due to heat transfer ($S_{\theta,\text{total}}$), fluid friction ($S_{\psi,\text{total}}$), combined total entropy generation ($S_{\text{total}}$), and average Bejan number ($\mathrm{Be}_{\text{av}}$) for power-law  indices $n = 0.6$, $1.0$, and $1.6$ in the square cavity under uniform and non-uniform bottom-wall heating.}
	\label{tab:Table_3}
	\begin{tabular}{c|cccc}
		\multicolumn{5}{c}{\textbf{Uniform heating bottom wall}} \\
		\hline
		$n$ & $S_{\theta,\text{total}}$ & $S_{\psi,\text{total}}$ & $S_{\text{total}}$ & $\mathrm{Be}_{\text{av}}$ \\
		\hline
		0.6 & 13.6196183 & 87.76998397 & 101.3896023 & 0.134329537 \\
		1.0 & 9.332793581 & 1.265645494 & 10.59843907 & 0.880581897 \\
		1.6 & 8.519506667 & 0.050049404 & 8.56955607  & 0.994159627 \\
	\end{tabular}
	
	\vspace{0.5cm}
	
	\begin{tabular}{c|cccc}
		\multicolumn{5}{c}{\textbf{Non-uniform heating bottom wall}} \\
		\hline
		$n$ & $S_{\theta,\text{total}}$ & $S_{\psi,\text{total}}$ & $S_{\text{total}}$ & $\mathrm{Be}_{\text{av}}$ \\
		\hline
		0.6 & 4.847500594 & 64.45955406 & 69.30705466 & 0.06942383 \\
		1.0 & 2.086104613 & 0.972145055 & 3.058249669 & 0.682123711 \\
		1.6 & 1.600679168 & 0.041713492 & 1.64239266  & 0.974601998 \\
	\end{tabular}
\end{table}

\begin{figure}
	\centering
	\includegraphics[width=1.0\textwidth]{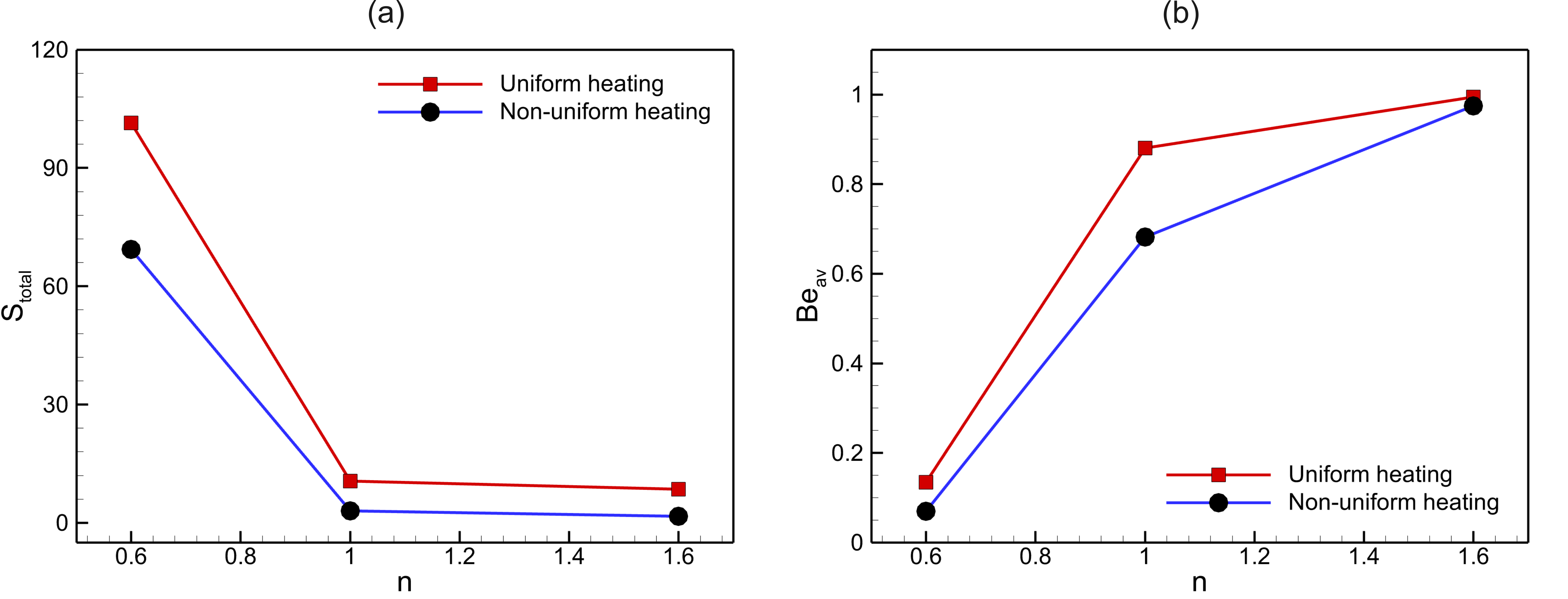}
	\caption{Variation of (a) total entropy generation $S_{\text{total}}$ 
		and (b) average Bejan number $\mathrm{Be}_{\text{av}}$ with power-law index $n$ 
		in the square cavity under uniform ($\theta = 1$) and non-uniform 
		($\theta = \sin(\pi X)$) bottom-wall heating.}
	\label{Fig:14}
\end{figure}

The total entropy generation due to heat transfer ($S_{\theta,\text{total}}$), fluid friction ($S_{\psi,\text{total}}$), 
combined total entropy generation ($S_{\text{total}}$), and average Bejan 
number ($\mathrm{Be}_{\text{av}}$) for both heating configurations are quantified in Table~\ref{tab:Table_3}. Under uniform heating, the shear-thinning fluid ($n = 0.6$) yields the highest $S_{\text{total}} = 101.39$, predominantly driven by viscous dissipation ($S_{\psi,\text{total}} = 87.77$), consistent with the intense and widespread 
$S_{\psi}$ contours observed near the cavity corners and bottom wall in 
Fig.~\ref{Fig:12}; the corresponding low $\mathrm{Be}_{\text{av}} = 0.134$ confirms 
that fluid friction irreversibility dominates overwhelmingly compared to heat 
transfer irreversibility for shear-thinning fluids under uniform heating. 
As $n$ increases to $1.6$, $S_{\psi,\text{total}}$ drops sharply to $0.050$, 
reducing $S_{\text{total}}$ to $8.57$, while $\mathrm{Be}_{\text{av}}$ approaches 
unity ($0.994$), indicating that heat transfer irreversibility becomes almost 
exclusively dominant - corroborated by the diffuse and weak $S_{\psi}$ 
contours and relatively persistent $S_{\theta}$ patterns seen in 
Fig.~\ref{Fig:10} for $n = 1.4$. Under non-uniform sinusoidal 
heating, the same rheological trend holds but with markedly reduced absolute 
values of both $S_{\theta,\text{total}}$ and $S_{\psi,\text{total}}$ across 
all $n$, reflecting the spatially localized nature of the sinusoidal thermal 
forcing, which concentrates entropy production near the cavity center rather 
than energizing the full bottom wall, as confirmed by the centrally confined 
$S_{\theta}$ and $S_{\psi}$ contours in Fig.~\ref{Fig:13}. Notably, for 
$n = 0.6$ under non-uniform heating, $S_{\text{total}}$ decreases from 
$101.39$ to $69.31$, with $\mathrm{Be}_{\text{av}}$ falling further to $0.069$, 
indicating an even stronger dominance of viscous dissipation compared to 
the uniform heating case. For $n = 1.0$ and $n = 1.6$, 
$Be_{\text{av}}$ increases toward unity under both heating conditions, as further illustrated by the monotonic decrease in $S_{\text{total}}$ and corresponding rise in $Be_{\text{av}}$ with increasing $n$ in Fig.~\ref{Fig:14}.

\begin{figure} 
	\centering
	\includegraphics[width=0.8\textwidth]{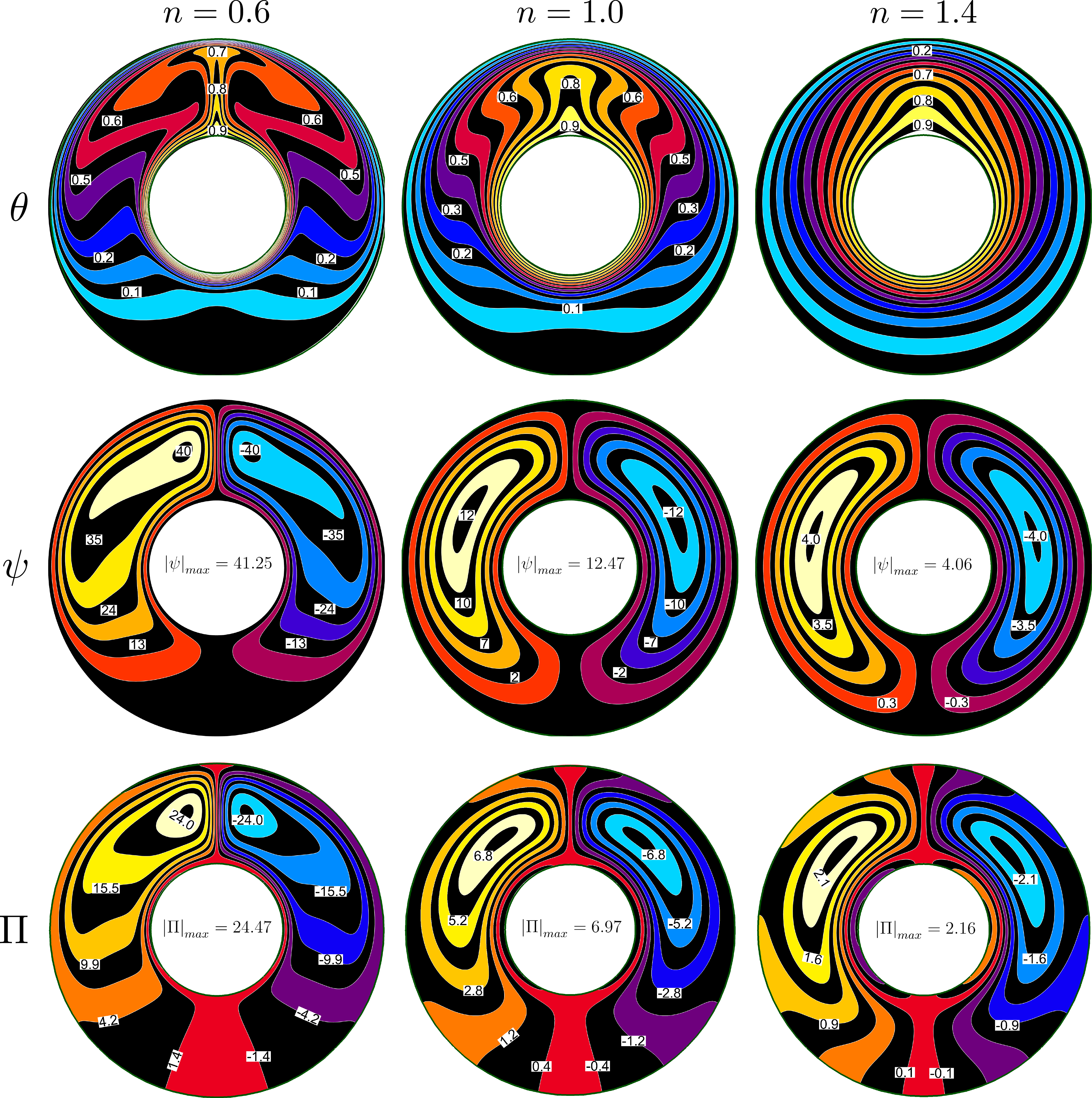}
	\caption{Isotherms $\theta$ (top), streamlines $\psi$ (middle), and 
		heatlines $\Pi$ (bottom) for power-law indices $n = 0.6$, $1.0$, and 
		$1.4$ in the concentric cylindrical annulus with uniform inner-wall 
		heating ($\theta = 1$).}
	\label{Fig:15}
\end{figure}

\subsection{Concentric cylindrical annulus configuration}
\label{Sec:6.3}

This subsection examines the effects of the power-law index $n$ and thermal 
boundary conditions on natural convection in the concentric cylindrical 
annulus, considering both uniform and non-uniform sinusoidal inner-wall 
heating at $\operatorname{Pr} = 100$ and $\operatorname{Ra} = 10^4$.

\subsubsection{Flow structure and thermal fields}
\label{Sec:6.3.1}

Figure~\ref{Fig:15} illustrates the isotherms $\theta$, streamlines $\psi$, 
and heatlines $\Pi$ for $n = 0.6$, $1.0$, and $1.4$ in the concentric 
cylindrical annulus under uniform inner-wall heating ($\theta = 1$). For $n = 0.6$, pronounced temperature gradients develop near the heated inner wall, and the isotherms are strongly distorted, reflecting vigorous buoyancy-driven convection. The flow is characterized by intense, asymmetric counter-rotating vortices with a maximum stream function magnitude of $|\psi|_{\max} = 41.25$, which significantly enhances convective heat transport. Correspondingly, the heatlines indicate strong advective heat transfer, with a high heat function magnitude ($\Pi_{\max} = 24.17$) concentrated near the inner wall and vortex cores, where steep thermal and velocity gradients prevail. As the fluid transitions to Newtonian behavior ($n = 1.0$), the flow structure becomes more symmetric and the isotherms gradually align, indicating a reduction in convective intensity. The circulation strength decreases substantially ($|\psi|_{\max} = 12.47$), accompanied by a marked reduction in heat transport ($\Pi_{\max} = 6.97$), signifying a shift toward moderately convection-dominated heat transfer. For $n = 1.4$, the increased effective viscosity suppresses fluid motion, resulting in weak circulation ($|\psi|_{\max} = 4.06$) and nearly concentric isotherms, indicative of conduction-dominated transport. The corresponding heatlines exhibit minimal distortion, and the reduced heat function magnitude ($\Pi_{\max} = 2.16$) reflects significantly lower thermal transport and irreversibility. 

\begin{figure} 
	\centering
	\includegraphics[width=0.8\textwidth]{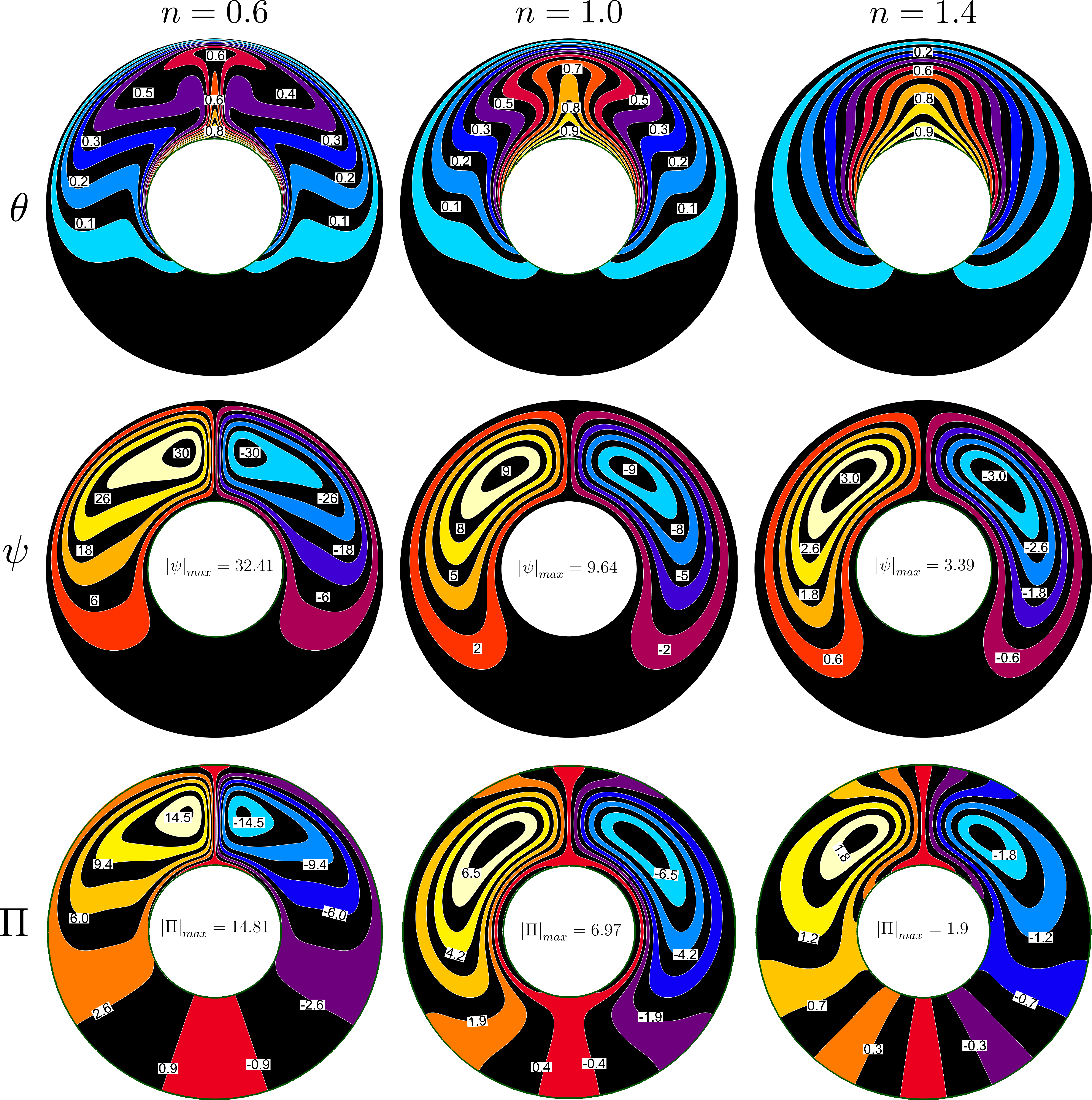}
	\caption{Isotherms $\theta$ (top), streamlines $\psi$ (middle), and 
		heatlines $\Pi$ (bottom) for power-law indices $n = 0.6$, $1.0$, and 
		$1.4$ in the concentric cylindrical annulus with non-uniform inner-wall 
		heating ($\theta = 0.5 (1+ \sin\varphi)$).}
	\label{Fig:16}
\end{figure}

Figure~\ref{Fig:16} presents the isotherms $\theta$, streamlines $\psi$, 
and heatlines $\Pi$ for $n = 0.6$, $1.0$, and $1.4$ in the concentric 
cylindrical annulus under non-uniform sinusoidal inner-wall heating 
($\theta = 0.5(1 + \sin\varphi)$). Compared to the uniform heating case 
(Fig.~\ref{Fig:15}), the spatially varying thermal boundary condition 
introduces pronounced azimuthal asymmetry in both the flow and thermal 
fields across all rheological regimes. For $n = 0.6$, the isotherms are 
strongly distorted and concentrated in the upper annular region where 
heating is maximum, driving asymmetric counter-rotating vortices with 
reduced circulation strength ($|\psi|_{\max} = 32.41$) compared to uniform 
heating ($|\psi|_{\max} = 41.25$); the heatlines confirm that convective 
heat transport is confined to preferred angular regions, yielding a lower 
heat function magnitude ($\Pi_{\max} = 14.81$). At $n = 1.0$, the flow 
weakens further ($|\psi|_{\max} = 9.64$) while retaining the azimuthal 
asymmetry imposed by the sinusoidal profile, and heat transport becomes 
more spatially confined ($\Pi_{\max} = 6.97$) relative to the uniform 
heating configuration. For $n = 1.4$, the increased apparent viscosity 
strongly suppresses fluid motion ($|\psi|_{\max} = 3.39$), producing 
nearly stratified isotherms and a significantly diminished heat function 
($\Pi_{\max} = 1.90$), indicating a conduction-dominated thermal field. 

\begin{figure} 
	\centering
	\includegraphics[width=0.8\textwidth]{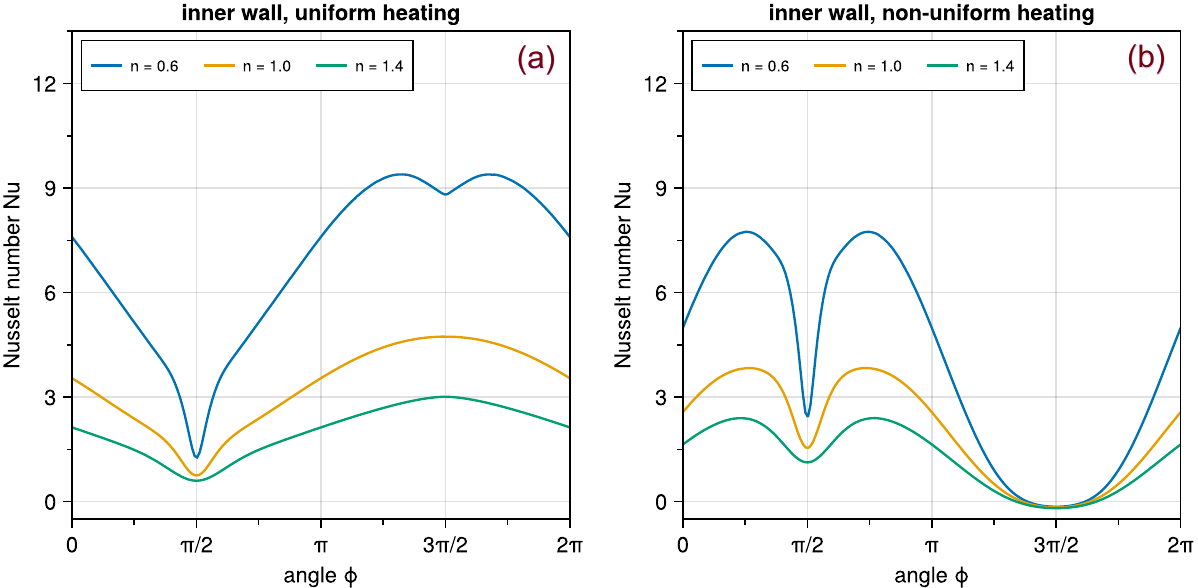}
	\caption{Local Nusselt number distributions along the inner wall of the 
		concentric cylindrical annulus for power-law indices $n = 0.6$, $1.0$, and $1.4$ under (a) uniform heating ($\theta = 1$) and (b) non-uniform sinusoidal heating ($\theta = 0.5 (1+ \sin\varphi)$).}
	\label{Fig:17}
\end{figure}
\begin{figure} 
	\centering
	\includegraphics[width=0.8\textwidth]{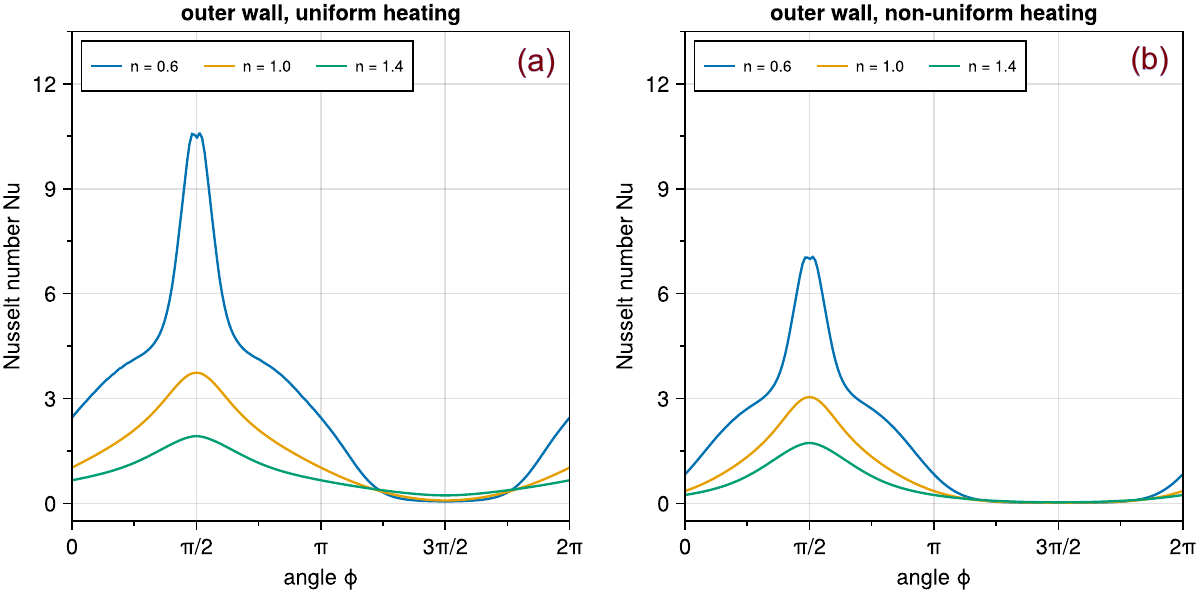}
	\caption{Local Nusselt number distributions along the outer wall of the 
		concentric cylindrical annulus for power-law indices $n = 0.6$, $1.0$, and $1.4$ under (a) uniform heating ($\theta = 1$) and (b) non-uniform sinusoidal heating ($\theta = 0.5 (1+ \sin\varphi)$).}
	\label{Fig:18}
\end{figure}

\subsubsection{Nusselt number and heat transfer}
\label{Sec:6.3.2}

Figure~\ref{Fig:17} presents the local Nusselt number distributions along 
the inner wall of the concentric cylindrical annulus for $n = 0.6$, $1.0$, 
and $1.4$ under uniform and non-uniform sinusoidal inner-wall 
heating. In both cases, the shear-thinning fluid ($n = 0.6$) yields the 
highest \textit{Nu} values owing to its reduced apparent viscosity, which 
intensifies buoyancy-driven circulation and thins the thermal boundary 
layer at the heated wall, while the shear-thickening fluid ($n = 1.4$) 
produces the lowest values due to viscous suppression of convective 
transport. Under uniform heating, the $\mathrm{Nu}$ distribution exhibits an 
asymmetric profile with elevated values over the upper annular region, 
where buoyancy-enhanced wall-normal heat transfer is strongest, and a 
pronounced minimum near $\varphi = \pi/2$, where local flow reorientation 
weakens thermal gradients at the wall. Under non-uniform sinusoidal 
heating, the $\mathrm{Nu}$ distribution develops a more structured angular variation 
with distinct peaks and troughs that follow the imposed thermal loading; 
the minimum near $\varphi \approx 3\pi/2$ corresponds to the weakly heated 
region where buoyancy production is smallest, while peak values in the 
upper sectors mark zones of stronger plume formation and more effective 
heat removal. Compared to uniform heating, non-uniform sinusoidal forcing 
redistributes heat transfer intensity into localized angular regions, with 
this localization most pronounced for $n = 0.6$, where the lower effective 
viscosity amplifies plume strength and angular non-uniformity in the 
thermal boundary layer.

Figure~\ref{Fig:18} presents the local Nusselt number distributions along 
the outer wall of the concentric cylindrical annulus for $n = 0.6$, $1.0$, 
and $1.4$ under uniform and non-uniform sinusoidal inner-wall 
heating. In both cases, the \textit{Nu} distribution exhibits a sharp peak near 
$\varphi = \pi/2$, corresponding to the uppermost point of the annulus 
where the buoyant thermal plume rising from the heated inner cylinder 
impinges most directly on the cold outer wall, generating the steepest 
wall-normal temperature gradients. The shear-thinning fluid ($n = 0.6$) 
produces the highest peak $\mathrm{Nu}$ value ($\mathrm{Nu}_{\max} \approx 10.5$ under 
uniform heating), as the reduced apparent viscosity intensifies plume-driven 
convection and enhances heat delivery to the outer wall, while the 
shear-thickening fluid ($n = 1.4$) yields a nearly flat, low-amplitude 
profile indicative of conduction-dominated transport with minimal convective 
contribution. Under non-uniform sinusoidal heating, the same sharp peak 
at $\varphi = \pi/2$ is preserved but with a noticeably reduced magnitude 
($\mathrm{Nu}_{\max} \approx 7.0$ for $n = 0.6$), since the spatially varying 
inner-wall temperature weakens the overall thermal driving force compared 
to uniform heating, while the near-zero $\mathrm{Nu}$ values in the lower annular 
region ($\varphi \approx \pi$ to $2\pi$) confirm that convective heat 
transfer to the outer wall is essentially absent where the inner-wall 
heating is minimal. Across both heating conditions, the rheological 
ordering ($n = 0.6 > n = 1.0 > n = 1.4$) is consistent, confirming that 
the power-law index governs the convective intensity at the outer wall 
irrespective of the thermal boundary condition imposed at the inner 
cylinder.

\begin{figure}
	\centering
	\includegraphics[width=0.8\textwidth]{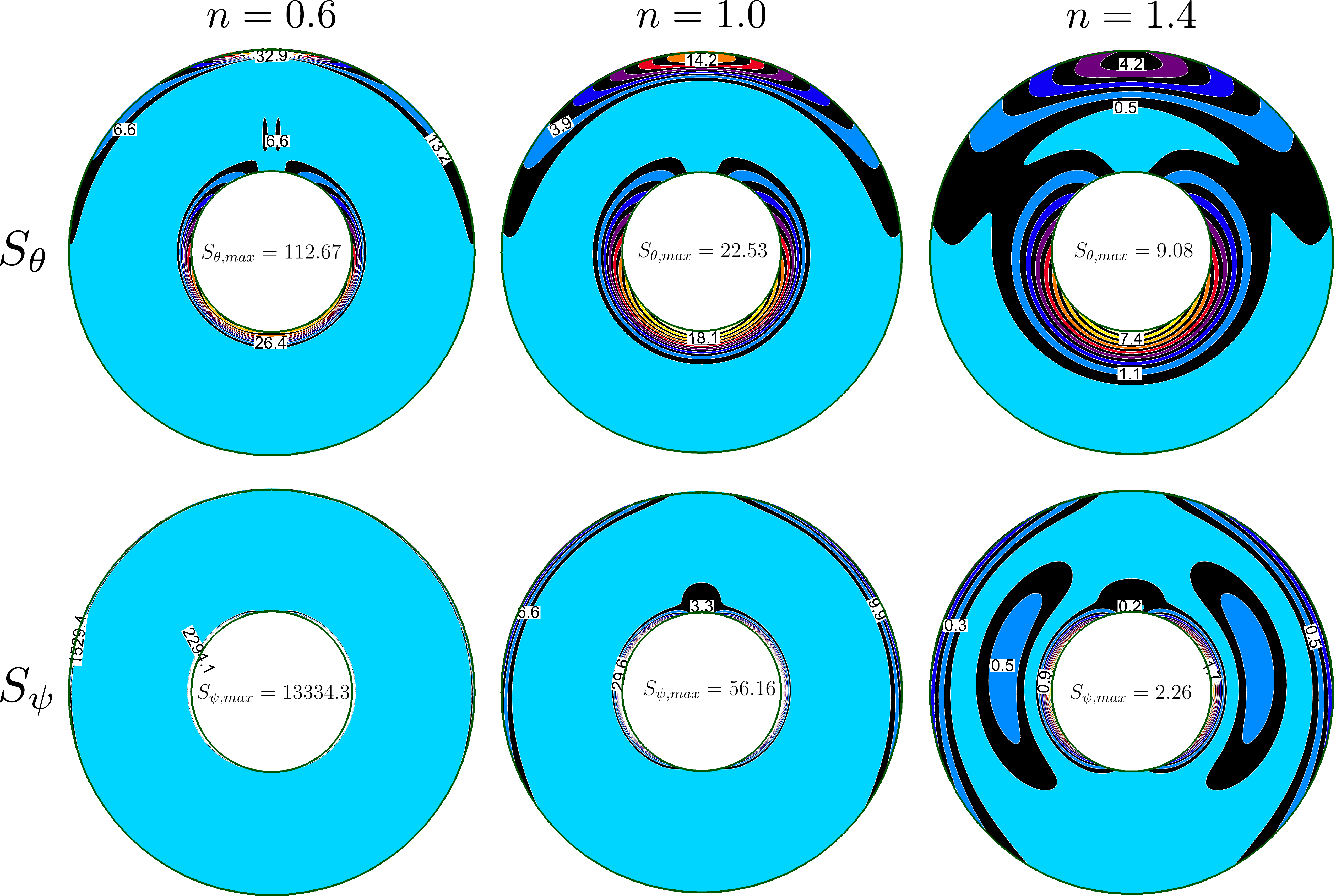}
	\caption{Contours of local entropy generation due to heat transfer 
		$S_{\theta}$ (top) and fluid friction $S_{\psi}$ (bottom) for 
		power-law indices $n = 0.6$, $1.0$, and $1.4$ in the concentric cylindrical annulus 
		with uniform bottom-wall heating ($\theta = 1$).}
	\label{Fig:19}
\end{figure}

\begin{figure}
	\centering
	\includegraphics[width=0.8\textwidth]{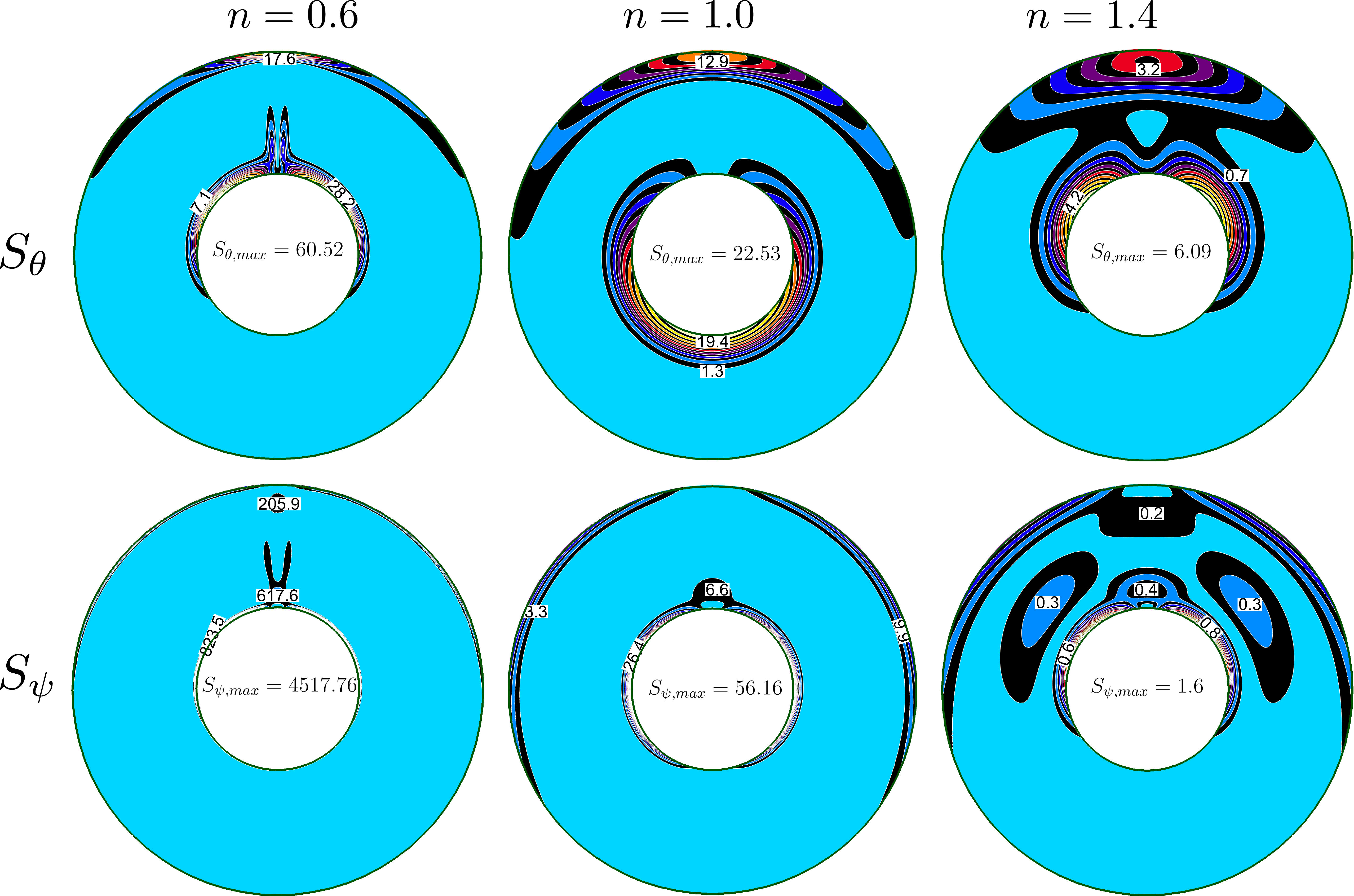}
	\caption{Contours of local entropy generation due to heat transfer	$S_{\theta}$ (top) and fluid friction $S_{\psi}$ (bottom) for 
		power-law indices $n = 0.6$, $1.0$, and $1.4$ in the concentric cylindrical annulus with non-uniform sinusoidal heating ($\theta = 0.5 (1+ \sin\varphi)$).}
	\label{Fig:20}
\end{figure}

\subsubsection{Entropy generation and Bejan number}
\label{Sec:6.3.3}

Figure~\ref{Fig:19} presents the contours of local entropy generation due 
to heat transfer $S_{\theta}$ and fluid friction $S_{\psi}$  for $n = 0.6$, $1.0$, and $1.4$ in the concentric cylindrical 
annulus under uniform inner-wall heating. For $n = 0.6$, $S_{\theta}$ 
attains its highest values near the inner heated wall ($S_{\theta,\max} = 
112.67$), where isothermal gradients are steepest as seen in 
Fig.~\ref{Fig:15}, while $S_{\psi}$ reaches extremely large magnitudes 
($S_{\psi,\max} = 13334.3$) spread across a significant portion of the 
annular domain, consistent with the intense circulation and strong velocity 
gradients of the shear-thinning flow, confirming viscous dissipation as 
the dominant source of irreversibility. At $n = 1.0$, both entropy 
components decrease substantially ($S_{\theta,\max} = 22.53$, $S_{\psi,\max} = 56.16$), and the contours become more localized near the inner wall, reflecting the reduced circulation strength ($|\psi|_{\max} = 12.47$) 
evident in Fig.~\ref{Fig:15}. For $n = 1.4$, entropy generation is further 
suppressed ($S_{\theta,\max} = 9.08$, $S_{\psi,\max} = 2.26$), with 
$S_{\theta}$ confined to a thin layer adjacent to the heated wall and 
$S_{\psi}$ nearly negligible throughout the domain, consistent with the 
weak velocity gradients and conduction-dominated thermal field observed 
in the streamlines and isotherms of Fig.~\ref{Fig:15}. 

Figure~\ref{Fig:20} presents the contours of local entropy generation due 
to heat transfer $S_{\theta}$ and viscous dissipation $S_{\psi}$ for $n = 0.6$, $1.0$, and $1.4$ in the concentric cylindrical 
annulus under non-uniform sinusoidal inner-wall heating ($\theta = 0.5 (1+ \sin\varphi)$). For $n = 0.6$, 
$S_{\theta}$ is highly localized in the upper annular region where the 
imposed temperature gradient is maximum, consistent with the asymmetric 
isotherm clustering observed in Fig.~\ref{Fig:16}, while $S_{\psi}$ 
remains the dominant irreversibility source; however, both its peak 
magnitude and spatial extent are considerably reduced compared to the 
uniform heating case (Fig.~\ref{Fig:19}), reflecting the confinement of 
vigorous convective motion to localized angular regions under the 
sinusoidal thermal forcing. At $n = 1.0$, both $S_{\theta}$ and $S_{\psi}$ 
decrease noticeably, with entropy contours concentrated near the heated 
inner wall, consistent with the weakened circulation ($|\psi|_{\max} = 
9.64$) in Fig.~\ref{Fig:16}; the reduction in $S_{\psi}$ is more 
pronounced than that in $S_{\theta}$ relative to Fig.~\ref{Fig:19}, 
highlighting the stronger suppression of large-scale convective motion 
under non-uniform heating. For $n = 1.4$, both entropy components are 
significantly diminished, with $S_{\theta}$ confined to a thin layer 
adjacent to the inner wall and $S_{\psi}$ appearing only in small isolated 
zones, consistent with the weak velocity gradients and near-conduction 
regime evident in Fig.~\ref{Fig:16}; the dominance of $S_{\theta}$ over 
$S_{\psi}$ at this $n$-value confirms a complete transition to 
heat-transfer-dominated irreversibility. 

\begin{table}[ht!]
	\centering
	\caption{Total entropy generation due to heat transfer ($S_{\theta,\text{total}}$), fluid friction ($S_{\psi,\text{total}}$), combined total entropy generation ($S_{\text{total}}$), and average Bejan number ($\mathrm{Be}_{\text{av}}$) for power-law  indices $n = 0.6$, $1.0$, and $1.6$ in the concentric cylinder annulus under uniform and non-uniform inner-wall heating.}
	\label{tab:Table_4}
	\vspace{0.3em}
	\begin{tabular}{ccccc}
		\multicolumn{5}{c}{\textbf{Uniform heating inner wall}} \\
		\toprule
		$n$ & $S_{\theta,\text{total}}$ & $S_{\psi,\text{total}}$ & $S_{\text{total}}$ & $\mathrm{Be}_{\text{av}}$ \\
		\midrule
		0.6 & 29.11686473 & 820.908914   & 850.025787   & 0.0342540961241055 \\
		1.0 & 13.89810836 & 15.43842255  & 29.33653091  & 0.473747506         \\
		1.6 & 8.514377316 & 1.452288412  & 9.966665728  & 0.85428543          \\
		\bottomrule
	\end{tabular}
	
	\vspace{0.5cm}
	
	\begin{tabular}{ccccc}
		\multicolumn{5}{c}{\textbf{Non-uniform heating inner wall}} \\
		\toprule
		$n$ & $S_{\theta,\text{total}}$ & $S_{\psi,\text{total}}$ & $S_{\text{total}}$ & $\mathrm{Be}_{\text{av}}$ \\
		\midrule
		0.6 & 12.31946233 & 280.5788535  & 292.8983158  & 0.042060543         \\
		1.0 & 6.093160385 & 6.603836072  & 12.69699646  & 0.479889902         \\
		1.6 & 3.711581468 & 0.683463739  & 4.395045207  & 0.844492216         \\
		\bottomrule
	\end{tabular}
\end{table}

Table~\ref{tab:Table_4} quantifies $S_{\theta,\text{total}}$, 
$S_{\psi,\text{total}}$, $S_{\text{total}}$, and $\mathrm{Be}_{\text{av}}$ for 
the concentric cylindrical annulus under both heating configurations. 
Under uniform heating, the shear-thinning fluid ($n = 0.6$) yields the 
highest $S_{\text{total}} = 850.03$, overwhelmingly dominated by viscous 
dissipation ($S_{\psi,\text{total}} = 820.91$, $\mathrm{Be}_{\text{av}} = 0.034$), 
consistent with the extremely large and spatially widespread $S_{\psi}$ 
contours observed in Fig.~\ref{Fig:19}, where intense circulation and 
strong velocity gradients drive irreversibility throughout the annular 
domain. As $n$ increases to $1.0$, $S_{\text{total}}$ 
drops sharply to $29.34$ with a more balanced contribution between 
$S_{\theta,\text{total}} = 13.90$ and $S_{\psi,\text{total}} = 15.44$ 
($\mathrm{Be}_{\text{av}} = 0.474$), reflecting the moderate and localized entropy 
contours in Fig.~\ref{Fig:19}. For $n = 1.6$, $S_{\psi,\text{total}}$ 
reduces drastically to $1.45$, driving $\mathrm{Be}_{\text{av}}$ to $0.854$ and 
confirming a near-complete transition to heat-transfer-dominated 
irreversibility, corroborated by the weak and confined $S_{\psi}$ 
contours in Fig.~\ref{Fig:19}. Under non-uniform heating, the same 
rheological trend persists but with substantially reduced absolute 
values across all $n$; notably, for $n = 0.6$, $S_{\text{total}}$ 
decreases from $850.03$ to $292.90$ and $\mathrm{Be}_{\text{av}}$ remains low 
($0.042$), confirming that viscous dissipation continues to dominate 
but is spatially confined to localized angular regions, as evidenced 
by the reduced $S_{\psi}$ contour extent in Fig.~\ref{Fig:20}. The near-identical $\mathrm{Be}_{\text{av}}$ values across both heating cases for 
$n = 1.0$ and $n = 1.6$ confirm that the power-law index, rather than the 
thermal boundary condition, governs the transition from viscous- to 
heat-transfer-dominated irreversibility in the annular configuration.

\begin{figure}
	\centering
	\includegraphics[width=1.0\textwidth]{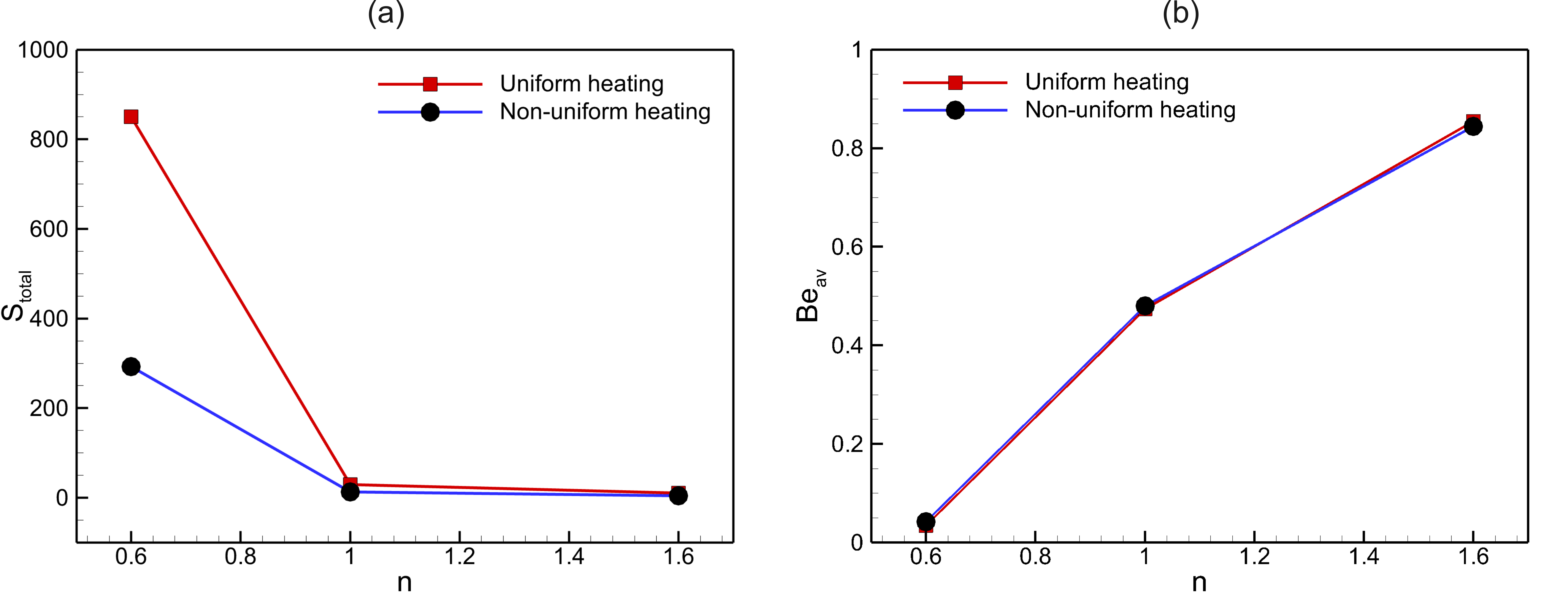}
	\caption{Variation of (a) total entropy generation $S_{\text{total}}$ 
		and (b) average Bejan number $Be_{\text{av}}$ with power-law index $n$ 
		in the concentric cylindrical annulus under uniform ($\theta = 1$) and non-uniform 
		($\theta = 0.5 (1+ \sin\varphi)$) inner wall heating.}
	\label{Fig:21}
\end{figure}

This trend is further quantified in Fig.~\ref{Fig:21}, where the total entropy generation $S_{\text{total}}$ decreases sharply with increasing power-law index $n$ for both heating conditions, with consistently lower values observed under non-uniform heating due to spatial localization of thermal forcing. Conversely, the average Bejan number $\mathrm{Be}_{\text{av}}$ increases monotonically with $n$, indicating a progressive shift from viscous-dominated irreversibility to heat-transfer-dominated behavior. Notably, the close agreement between uniform and non-uniform heating curves for $\mathrm{Be}_{\text{av}}$ suggests that the rheological effects dominate the irreversibility mechanism, while thermal boundary conditions primarily influence its magnitude rather than its nature.

\section{Concluding remarks and outlook}\label{Sec:7}

This study examined the role of thermal boundary conditions on natural convection and entropy generation in non-Newtonian power-law fluids confined within a square cavity and a concentric cylindrical annulus. A \texttt{Gridap.jl} finite element 
framework was developed for the steady two-dimensional governing equations under the Boussinesq approximation and validated against established benchmark solutions for both Newtonian and non-Newtonian cases. The good agreement obtained for isotherm fields, streamlines, local Nusselt number distributions, and entropy generation contours confirms the accuracy and robustness of the present numerical methodology.

The results demonstrate that thermal boundary conditions strongly influence the structure and intensity of buoyancy-driven transport, while their interaction with fluid rheology governs the resulting thermodynamic irreversibility. Uniform heating promotes stronger and more spatially distributed circulation, yielding higher heat transfer rates and larger total entropy generation. In contrast, non-uniform sinusoidal heating localizes the thermal forcing, confines plume development to restricted regions, and consistently reduces the total entropy generation in both geometries. At the same time, the power-law index remains a key controlling parameter: shear-thinning fluids enhance convection and intensify thermal and velocity gradients, whereas shear-thickening fluids suppress circulation and shift the system toward conduction-dominated behavior.

Entropy generation analysis further shows that the dominant source of irreversibility depends primarily on the rheological nature of the fluid, whereas the imposed thermal boundary condition determines its magnitude and spatial distribution. In shear-thinning fluids, viscous dissipation contributes significantly to the total irreversibility, particularly in the annular geometry where circulation is strongest. As the power-law index increases, the contribution of fluid-friction irreversibility decreases rapidly, and the Bejan number rises, indicating a transition toward heat-transfer-dominated entropy generation. The findings highlight that thermal boundary design offers an effective means of controlling heat transfer performance and thermodynamic losses in systems involving power-law fluids. Future work may extend the present framework to three-dimensional flows, higher Rayleigh number regimes, and more complex constitutive models, as well as to optimization strategies for thermally efficient systems with non-Newtonian working fluids.

\section*{Acknowledgments}
L.T. acknowledges the support by the German Research Foundation (DFG), project B04 (504291427) within the SFB 1481 (442047500).
S.S. acknowledge the support by the German Research Foundation (DFG), within the Research Unit FOR5409 under project number 463312734. 

\section*{Data availability \& reproducibility}
We provide the source code, scripts, and data to reproduce the simulations results on GitHub (\url{https://github.com/lamBOOO/GeMotion.jl}) and Zenodo~\cite{theisen2026gemotion}.


\appendix
\section{Appendix: Nomenclature} \label{Appendix:A}

\begin{tabular}{ll}
	$u, v$                 & Velocity components in $x$ and $y$ directions (m/s) \\
	$U, V$                 & Dimensionless velocity components \\
	$x, y$                 & Cartesian coordinates (m) \\
	$X, Y$                 & Dimensionless coordinates \\
	$r, \varphi$           & Cylindrical radial and angular coordinates (m, rad) \\
	$p$                    & Pressure (Pa) \\
	$P$                    & Dimensionless pressure \\
	$T$                    & Temperature (K) \\
	$\theta$               & Dimensionless temperature \\
	$T_h$                  & Hot wall temperature (K) \\
	$T_c$                  & Cold wall temperature (K) \\
	$\psi$                 & Stream function \\
	$\Pi$                  & Heat function \\
	$\Phi_i$               & Basis function in finite element formulation \\
	$\tau_{ij}$            & Shear stress tensor components (Pa) \\
	$\mu_a$                & Apparent viscosity (Pa$\cdot$s) \\
	$\alpha$               & Thermal diffusivity (m$^2$/s) \\
	$\nu$                  & Kinematic viscosity (m$^2$/s) \\
	$K$                    & Consistency index in power-law fluid (Pa$\cdot$s$^n$) \\
	$n$                    & Power-law index \\
	$\rho$                 & Fluid density (kg/m$^3$) \\
	$\beta$                & Thermal expansion coefficient (1/K) \\
	$g$                    & Gravitational acceleration (m/s$^2$) \\
	$\widehat{\boldsymbol{g}}$ & Dimensionless buoyancy direction vector \\
	$L$                    & Characteristic length (m) \\
	$R_i$                  & Inner cylinder radius (m) \\
	$R_o$                  & Outer cylinder radius (m) \\
	$\text{Ra}$            & Rayleigh number \\
	$\text{Pr}$            & Prandtl number \\
	$\text{Nu}$            & Nusselt number \\
	$\text{Nu}_b$, $\text{Nu}_s$ & Local Nusselt number at bottom and side walls \\
	$\text{Nu}_{avg}$      & Average Nusselt number \\
	$S_\theta$             & Local entropy generation due to heat transfer \\
	$S_\psi$               & Local entropy generation due to fluid friction \\
	$S_{\theta,\text{total}}$ & Total entropy generation from heat transfer \\
	$S_{\psi,\text{total}}$   & Total entropy generation from fluid friction \\
	$S_{\text{total}}$     & Total entropy generation \\
	$\text{Be}_{av}$       & Average Bejan number \\
	$\phi$                 & Irreversibility distribution parameter \\
\end{tabular}

\section{Appendix: Variational formulation and finite element discretization}\label{Appendix:B}

The variational formulation is derived in a general form so that it can be
applied to the square cavity, the annular domain, and more general
two-dimensional geometries \(\Omega \subset \mathbb{R}^2\). The boundary
\(\Gamma = \partial \Omega\) is decomposed as
\(\Gamma = \Gamma_{\mathrm{D}} \cup \Gamma_{\mathrm{N}}\), with
\(\Gamma_{\mathrm{D}} \cap \Gamma_{\mathrm{N}} = \emptyset\), in order to
accommodate Dirichlet and Neumann boundary conditions in a unified manner.
For the derivation, a vector notation is adopted by introducing
\(\boldsymbol{\mathcal{U}} \coloneq (\boldsymbol{U}, P, \theta)
\coloneq ((U,V), P, \theta)\) as the vector of unknown fields. Furthermore, the gradient operator is denoted by
\(\boldsymbol{\nabla} \coloneq (\partial_X,\partial_Y)^{\top}\), and the
divergence of the velocity field is written as
\(\boldsymbol{\nabla} \cdot \boldsymbol{U} = \partial_X U + \partial_Y V\).
The outward unit normal vector on the boundary is denoted by
\(\boldsymbol{n}\).

The steady, dimensionless governing equations for incompressible natural
convection of a power-law fluid from~\eqref{eq:10}--\eqref{eq:13} can then be
written as
\begin{equation}\label{eq_incompressible_navierstokes}
	\left\{
	\begin{aligned}
		\boldsymbol{\nabla} \cdot \boldsymbol{U} &= 0
		\textnormal{ in } \Omega
		,
		\\
		(\boldsymbol{U} \cdot \boldsymbol{\nabla}) \boldsymbol{U} + \boldsymbol{\nabla} P - 2\operatorname{Pr} \boldsymbol{\nabla} \cdot \left(\bar{\mu}_{\textnormal{s}}(\boldsymbol{\nabla} \boldsymbol{U}) D(\boldsymbol{\nabla} \boldsymbol{U})\right) - \operatorname{Ra}\operatorname{Pr} \theta \widehat{\boldsymbol{g}} &= \boldsymbol{0} \textnormal{ in } \Omega
		,
		\\
		\boldsymbol{U} \cdot \boldsymbol{\nabla} \theta - \boldsymbol{\nabla} \cdot (\boldsymbol{\nabla } \theta) &= 0 \textnormal{ in } \Omega
		,
		\\
		\boldsymbol{U} = \boldsymbol{0} \textnormal{ on } \Gamma
		, \;
		\boldsymbol{\nabla} \theta \cdot \boldsymbol{n} = 0 \textnormal{ on } \Gamma_{\textnormal{N}}
		, \;
		\theta &= \theta_{\mathrm{D}} \textnormal{ on } \Gamma_{\textnormal{D}}
		.
	\end{aligned}
	\right.
\end{equation}
In Eq.~\eqref{eq_incompressible_navierstokes}, \(\boldsymbol{U}=(U,V)\),
\(P\), and \(\theta\) denote the dimensionless velocity, pressure, and
temperature fields, respectively. No-slip and
impermeability conditions are imposed on the entire boundary \(\Gamma\),
whereas the thermal boundary conditions are prescribed through a mixed
Dirichlet--Neumann description: \(\theta=\theta_{\mathrm{D}}\) on \(\Gamma_{\mathrm{D}}\) and
\(\boldsymbol{\nabla}\theta\cdot\boldsymbol{n}=0\) on the adiabatic part
\(\Gamma_{\mathrm{N}}\).
The vector \(\widehat{\boldsymbol{g}}\) denotes the dimensionless buoyancy 
direction defined in Sec.~\ref{Sec:2.4}.
To avoid singularities when \(|\boldsymbol{\nabla} \boldsymbol{U}| \to 0\), we stabilize the dimensionless apparent viscosity by introducing a small regularization parameter \(c_{\textnormal{stab}} > 0\) and use
\begin{equation}
	\bar{\mu}_{\textnormal{s}}(\boldsymbol{\nabla}\boldsymbol{U}) =
	\left(
	c_{\textnormal{stab}}
	+ 2 D(\boldsymbol{\nabla} \boldsymbol{U}) : D(\boldsymbol{\nabla} \boldsymbol{U})
	\right)^{\frac{n-1}{2}},
	\qquad c_{\textnormal{stab}} = 10^{-3}.
\end{equation}

The trial space is denoted by \(\mathbb{U}\) and includes the prescribed 
velocity and temperature Dirichlet data. The corresponding test space is 
denoted by \(\mathbb{V}\), with homogeneous conditions on the Dirichlet 
boundaries; the pressure component is constrained to have zero mean. For 
each \(\boldsymbol{\mathcal{U}}\in\mathbb{U}\), we introduce a test 
function \(\boldsymbol{\mathcal{V}} = (\boldsymbol{v}, q, \zeta)\in
\mathbb{V}\). We then multiply~\eqref{eq_incompressible_navierstokes} by 
their respective test functions and integrate over \(\Omega\). Integration 
by parts and summing all equations leads to the following nonlinear 
variational problem: Find \(\boldsymbol{\mathcal{U}} = (\boldsymbol{U}, P, \theta)\in\mathbb{U}\), such that for all \(\boldsymbol{\mathcal{V}} = (\boldsymbol{v}, q, \zeta) \in \mathbb{V}\):
\begin{equation}\label{eq_nonlinear_weak_form}
  R((\boldsymbol{U},P,\theta),(\boldsymbol{v},q,\zeta))
  =
  a((\boldsymbol{U},P,\theta),(\boldsymbol{v},q,\zeta))
  + b_{\boldsymbol{U}}(\boldsymbol{v})
  + c_{\boldsymbol{U}}(\boldsymbol{v})
  + d_{\boldsymbol{U},\theta}(\zeta)
  =
  0
  ,
\end{equation}
with the bilinear form
\begin{equation}
\begin{aligned}
  a((\boldsymbol{U},P,\theta),(\boldsymbol{v},q,\zeta))
  =
  - \int_\Omega (\boldsymbol{\nabla} \cdot \boldsymbol{v}) P \,\textnormal{d}\Omega
  + \int_\Omega (\boldsymbol{\nabla} \cdot \boldsymbol{U}) q \,\textnormal{d}\Omega + \int_\Omega \boldsymbol{\nabla} \theta \cdot \boldsymbol{\nabla} \zeta \,\textnormal{d}\Omega
  - \operatorname{Ra}\operatorname{Pr} \int_\Omega \theta \widehat{\boldsymbol{g}} \cdot \boldsymbol{v} \,\textnormal{d}\Omega
  ,
\end{aligned}
\end{equation}
and the \(\boldsymbol{\mathcal{U}}\)-dependent semilinear forms for 
viscous diffusion, convective momentum transport, and energy transport,
\begin{align}
  b_{\boldsymbol{U}}(\boldsymbol{v}) =
  2 \operatorname{Pr} \int_\Omega \bar{\mu}_{\textnormal{s}}(\boldsymbol{\nabla} \boldsymbol{U}) \left(D(\boldsymbol{\nabla} \boldsymbol{v}) \boldsymbol{:} D(\boldsymbol{\nabla} \boldsymbol{U})\right) \,\textnormal{d}\Omega
  , \;\;
  c_{\boldsymbol{U}}(\boldsymbol{v}) = \int_\Omega ((\boldsymbol{U} \cdot \boldsymbol{\nabla}) \boldsymbol{U}) \cdot \boldsymbol{v} \,\textnormal{d}\Omega
  , \;\;
  d_{\boldsymbol{U},\theta}(\zeta) = \int_\Omega (\boldsymbol{U} \cdot \boldsymbol{\nabla} \theta) \zeta \,\textnormal{d}\Omega
  .
\end{align}
We note here, that for the Newtonian case, i.e., \(n=1\), the form \(b(\boldsymbol{U},\boldsymbol{v}) \coloneq b_{\boldsymbol{U}}(\boldsymbol{v})\) is bilinear and resembles the classical viscous term in the Navier--Stokes equations since the apparent viscosity is constant.

\par
For the nonlinear residual \(R(\boldsymbol{\mathcal{U}},\boldsymbol{\mathcal{V}})\) from the weak form~\eqref{eq_nonlinear_weak_form}, given a starting point \(\boldsymbol{\mathcal{U}}_{0}\), we write \(\boldsymbol{\mathcal{U}}_{k+1} = \boldsymbol{\mathcal{U}}_{k} + \delta \boldsymbol{\mathcal{U}}_{k}\)~\cite[p324ff]{elmanFiniteElementsFast2005}. The linearization of \(R\) yields the Jacobian
\begin{equation}
  R_{\boldsymbol{\mathcal{U}}_k}'(\delta\boldsymbol{\mathcal{U}}_k,\boldsymbol{\mathcal{V}})
  =
  a(\delta\boldsymbol{\mathcal{U}}_k,\boldsymbol{\mathcal{V}})
  +
  \textnormal{d}b(\boldsymbol{U}_k, \delta\boldsymbol{U}_k, \boldsymbol{v})
  +
  \textnormal{d}c(\boldsymbol{U}_k, \delta\boldsymbol{U}_k, \boldsymbol{v})
  +
  \textnormal{d}d(\boldsymbol{U}_k, \delta\boldsymbol{U}_k, \theta_k, \delta \theta_k, \zeta)
  .
\end{equation}
The viscous contribution to the Jacobian involves the derivative of the power-law viscosity:
\begin{equation}
  \textnormal{d}b(\boldsymbol{U}_k, \delta\boldsymbol{U}_k, \boldsymbol{v}) = \int_\Omega 2 \operatorname{Pr} \left[ \textnormal{d}\bar{\mu}_{\textnormal{s}}(\boldsymbol{\nabla}\boldsymbol{U}_k;\boldsymbol{\nabla}\delta\boldsymbol{U}_k) D(\boldsymbol{\nabla} \boldsymbol{U}_k) + \bar{\mu}_{\textnormal{s}}(\boldsymbol{\nabla}\boldsymbol{U}_k) D(\boldsymbol{\nabla}\delta\boldsymbol{U}_k) \right] : D(\boldsymbol{\nabla} \boldsymbol{v}) \,\textnormal{d}\Omega
  .
\end{equation}
To derive the explicit expression for \(\textnormal{d}\bar{\mu}_{\textnormal{s}}\), we apply the chain rule to the stabilized apparent viscosity definition. Let \(\gamma(\boldsymbol{\nabla}\boldsymbol{U}) = c_{\textnormal{stab}} + 2[D(\boldsymbol{\nabla} \boldsymbol{U}) : D(\boldsymbol{\nabla} \boldsymbol{U})]\) be the regularized shear-rate magnitude. Then, in the direction \(\boldsymbol{\nabla}\delta\boldsymbol{U}\), we have
\begin{equation}
	\textnormal{d}\bar{\mu}_{\textnormal{s}}(\boldsymbol{\nabla}\boldsymbol{U};\boldsymbol{\nabla}\delta\boldsymbol{U})
	=
	\frac{n-1}{2}
	\gamma(\boldsymbol{\nabla}\boldsymbol{U})^{\frac{n-3}{2}}
	4\left(D(\boldsymbol{\nabla}\boldsymbol{U}) : D(\boldsymbol{\nabla}\delta\boldsymbol{U})\right)
	.
\end{equation}
This derivative accounts for the nonlinear dependence of the apparent viscosity on the velocity gradient, which is essential for the Newton-Raphson convergence. For \(n<1\), the negative exponent in the regularized shear-rate invariant makes the viscosity derivative especially sensitive in low-shear regions. For \(n=1\), the derivative vanishes and the Newtonian case reduces to a constant-viscosity contribution.
\par
The convective terms contribute with
\begin{equation}
  \textnormal{d}c(\boldsymbol{U}_k, \delta\boldsymbol{U}_k, \boldsymbol{v})
  =
  \int_\Omega ((\boldsymbol{U}_k \cdot \boldsymbol{\nabla}) \delta\boldsymbol{U}_k) \cdot \boldsymbol{v} \,\textnormal{d}\Omega
  +
  \int_\Omega ((\delta\boldsymbol{U}_k \cdot \boldsymbol{\nabla}) \boldsymbol{U}_k) \cdot \boldsymbol{v} \,\textnormal{d}\Omega
  ,
\end{equation}
while the thermal advection contribution is
\begin{equation}
  \textnormal{d}d(\boldsymbol{U}_k, \delta\boldsymbol{U}_k, \theta_k, \delta \theta_k, \zeta)
  =
  \int_\Omega (\delta\boldsymbol{U}_k \cdot \boldsymbol{\nabla} \theta_k) \zeta \,\textnormal{d}\Omega
  +
  \int_\Omega (\boldsymbol{U}_k \cdot \boldsymbol{\nabla} \delta \theta_k) \zeta \,\textnormal{d}\Omega
  .
\end{equation}
The resulting Newton update rule per iteration is given by the solution to the following linear system: Find \(\delta \boldsymbol{\mathcal{U}}_k\), such that
\begin{equation}
  R_{\boldsymbol{\mathcal{U}}_k}'(\delta\boldsymbol{\mathcal{U}}_k,\boldsymbol{\mathcal{V}})
  =
  -
  R(\boldsymbol{\mathcal{U}}_k,\boldsymbol{\mathcal{V}}) \quad \forall \boldsymbol{\mathcal{V}} \in \mathbb{V}
\end{equation}
\par
In the discrete setting, this translates to solving the linear system \(\mathbf{J}^{(k)} \delta\mathbf{U}_h^{(k)} = -\mathbf{R}^{(k)}\), where \(\mathbf{J}^{(k)}\) is the assembled Jacobian matrix, \(\delta\mathbf{U}_h^{(k)}\) is the vector of degrees of freedom for the Newton update, and \(\mathbf{R}^{(k)}\) is the residual vector. After solving for \(\delta\mathbf{U}_h^{(k)}\), the new iterate is computed as \(\mathbf{U}_h^{(k+1)} = \mathbf{U}_h^{(k)} + \delta\mathbf{U}_h^{(k)}\).

\printcredits{}

\apptocmd{\sloppy}{\hbadness10000\relax}{}{}




\end{document}